\documentclass[%
reprint,
superscriptaddress,
footinbib,
twocolumn,
amsmath,amssymb,
aps,
prl,
floatfix,
longbibliography
]{revtex4-2}

\usepackage{graphicx}
\usepackage{bm}

\usepackage[colorlinks=true, linkcolor=blue,anchorcolor=blue, citecolor=blue,urlcolor=blue]{hyperref}

\usepackage[normalem]{ulem}
\usepackage{braket}
\usepackage{multirow}
\begin{document}
\preprint{APS/123-QED}

\title{Long-Range Spin-Orbit-Coupled Magnetoelectricity in Type-II Multiferroic NiI$_2$}

\author{Weiyi Pan}
\altaffiliation{Contributed equally to this work.}
\affiliation{State Key Laboratory of Low Dimensional Quantum Physics and Department of Physics, Tsinghua University, Beijing 100084, China}

\author{Zefeng Chen}
\altaffiliation{Contributed equally to this work.}
\affiliation{Key Laboratory of Computational Physical Sciences (Ministry of Education), Institute of Computational Physical Sciences, State Key Laboratory of Surface Physics, and Department of Physics, Fudan University, Shanghai 200433, China.}

\author{Dezhao Wu}
\affiliation{State Key Laboratory of Low Dimensional Quantum Physics and Department of Physics, Tsinghua University, Beijing 100084, China}
\author{Weiqin Zhu}
\affiliation{Key Laboratory of Computational Physical Sciences (Ministry of Education), Institute of Computational Physical Sciences, State Key Laboratory of Surface Physics, and Department of Physics, Fudan University, Shanghai 200433, China.}
\author{Zhiming Xu}
\affiliation{State Key Laboratory of Low Dimensional Quantum Physics and Department of Physics, Tsinghua University, Beijing 100084, China}
\author{Lianchuang Li}
\affiliation{Key Laboratory of Computational Physical Sciences (Ministry of Education), Institute of Computational Physical Sciences, State Key Laboratory of Surface Physics, and Department of Physics, Fudan University, Shanghai 200433, China.}

\author{Junsheng Feng}
\affiliation{School of Physics and Materials Engineering, Hefei Normal University, Hefei 230601, China}

\author{Bing-Lin Gu}
\affiliation{State Key Laboratory of Low Dimensional Quantum Physics and Department of Physics, Tsinghua University, Beijing 100084, China}
\affiliation{Institute for Advanced Study, Tsinghua University, Beijing 100084, China}

\author{Wenhui Duan}
\email{duanw@tsinghua.edu.cn}
\affiliation{State Key Laboratory of Low Dimensional Quantum Physics and Department of Physics, Tsinghua University, Beijing 100084, China}
\affiliation{Institute for Advanced Study, Tsinghua University, Beijing 100084, China}
\affiliation{Frontier Science Center for Quantum Information, Beijing 100084, China}

\author{Changsong Xu}
\email{csxu@fudan.edu.cn}
\affiliation{Key Laboratory of Computational Physical Sciences (Ministry of Education), Institute of Computational Physical Sciences, State Key Laboratory of Surface Physics, and Department of Physics, Fudan University, Shanghai 200433, China.}
\affiliation{Hefei National Laboratory, Hefei 230088, China}

\begin{abstract}
Type-II multiferroics, where spin order induces ferroelectricity, exhibit strong magnetoelectric coupling. However, for the typical 2D type-II multiferroic NiI$_2$, the underlying magnetoelectric mechanism remains unclear. Here, applying generalized spin-current model, together with first-principles calculations and a tight-binding approach, we build a comprehensive magnetoelectric model for spin-induced polarization. 
Such model reveals that the spin-orbit coupling extends its influence to the third-nearest neighbors, whose contribution to polarization  rivals that of the first-nearest neighbors. 
By analyzing the orbital-resolved contributions to polarization, our tight-binding model reveals that the long-range magnetoelectric coupling is enabled by the strong $e_g$-$p$ hopping of NiI$_2$. Monte Carlo simulations further predict a Bloch-type magnetic skyrmion lattice at moderate magnetic fields, accompanied by polar vortex arrays. These findings can guide the discovery and design of strongly magnetoelectric multiferroics.
\end{abstract}

\maketitle

\begin{figure}[t]
\includegraphics[width=8.5cm]{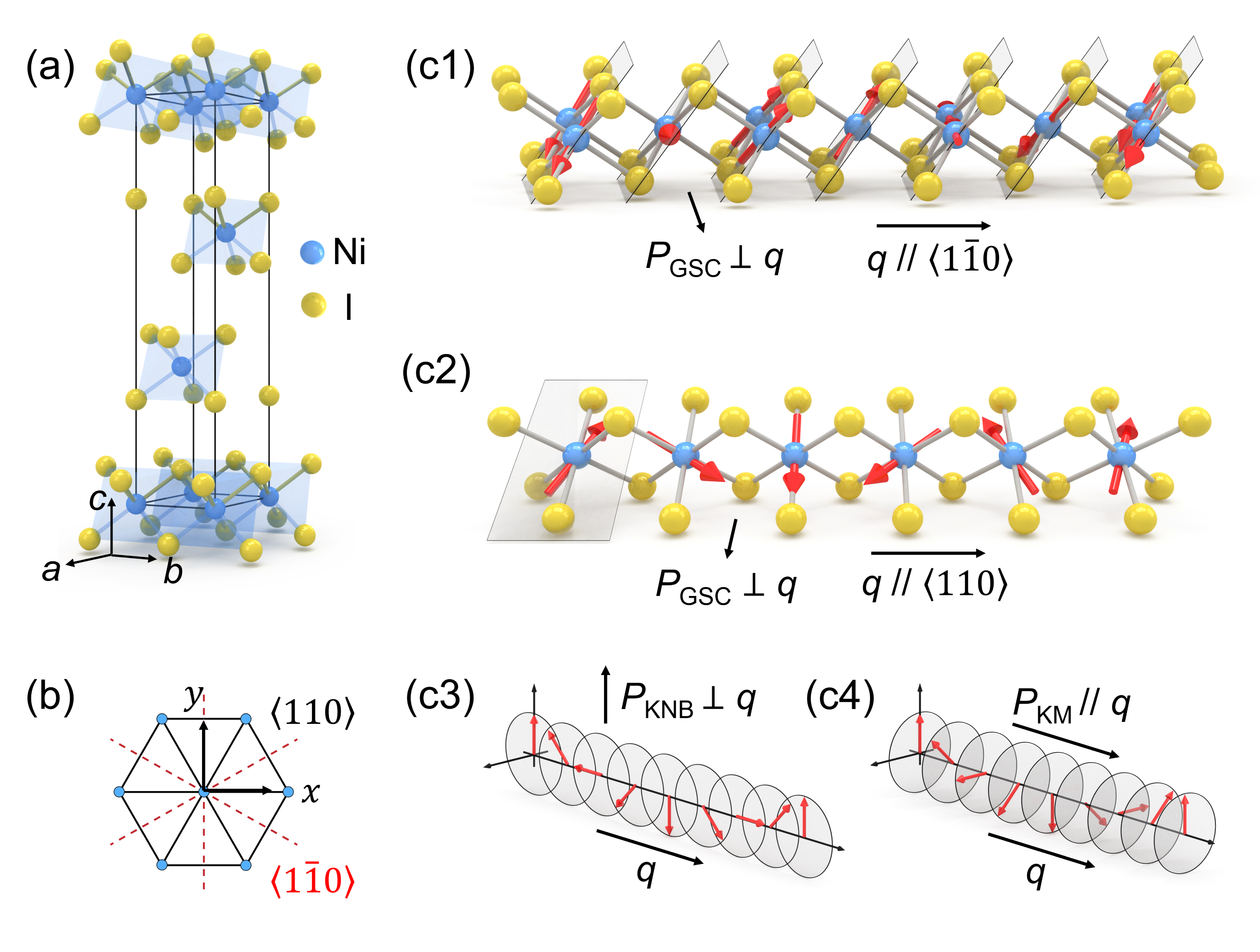}
\caption{Schematics of (a) crystal structure of NiI$_{2}$  and (b) its top view highlighting key orientations. Panel (c) illustrates four representative noncollinear spin states:  (c1) canted screw propagating along $\langle 1\overline{1}0\rangle$, (c2) canted cycloid propagating along $\langle 110\rangle$, (c3) a vertical cycloid, and (c4) a proper screw. 
The half-transparent planes in  (c) indicate the spin rotation planes, while the black arrows denote the propagation vector $\mathbf{q}$ and the spin-induced polarization $\mathbf{P}$. } 

\label{1}
\end{figure}

Type-II multiferroicity arises when electric polarization is induced by magnetic order, typically leading to strong magnetoelectric coupling \cite{MF2,MF3,MF4,XCS}. NiI$_2$ is a typical van der Waals type-II multiferroic that crystallizes in $D_{3d}$ symmetry, featuring triangular-latticed layers of Ni$^{2+}$ ions [Fig. \ref{1}(a)]. Each Ni$^{2+}$ (3$d^8$, $S=1$) carries a local moment of about 2\,$\mu_B$, arranged in a canted screw state propagating along $\langle 1\bar{1}0\rangle$ directions in the layer [Fig. \ref{1}(b, c1)] \cite{kuindersma1981magnetic,kurumaji2013magnetoelectric}. 
A realistic spin model has been developed and demonstrates that the screw originates from competitions of Heisenberg terms and  the canting of this screw results from a novel Kitaev interaction \cite{NiI2Xu}.
The canted screw induces an in-plane polarization perpendicular to its propagation, giving rise to type-II multiferroicity \cite{kurumaji2013magnetoelectric}. Recently, NiI$_2$ was successfully fabricated in few-layer or monolayer forms, 
where the helical state shifts to propagate along $\langle 110 \rangle$ directions \cite{NiI2Fu, NiI2AM,NiI2Nature,NiI2Nano}, which gives rise to distinct spin rotation plane, as well as possible parallel or perpendicular polarizations [Fig. \ref{1}(c2)] \cite{NiI2Nature, NiI2AM,NiI2Nano}.


Despite the established multiferroic behaviors and the realistic spin model, the magnetoelectric coupling mechanism of NiI$_2$ remains elusive. 
From a symmetry perspective, the Ni$^{2+}$ ions occupy inversion centers, thus ruling out both $p$-$d$ hybridization (where the polarization is induced by the charge transfer between metal and ligand \cite{arima2007ferroelectricity,jia2006bond,zhu2024mechanism}) and the anisotropic symmetric exchange mechanism \cite{feng2016anisotropic}.  Moreover, spin-orbit coupling (SOC) has been shown to be essential for generating polarization in NiI$_2$ \cite{KNB2}, which excludes the exchange striction mechanism, since it induces polarization without relying on relativistic effects \cite{sergienko2006ferroelectricity,picozzi2007dual}. 
In an effort to describe NiI$_2$ multiferroicity,  
 the KNB (spin-current) model was applied \cite{KNB3,antao2024electric}, where the local electric dipole is related to spins via $\mathbf{P}_{\rm KNB} \propto \bm{e}_{ij} \times (\mathbf{S}_i \times \mathbf{S}_j)$, with $\mathbf{e}_{ij}$ being the vector from $\mathbf{S}_i$ to $\mathbf{S}_j$ \cite{KNB} [Fig. \ref{1}(c3)]. Although this model works well for terbium manganite by reproducing the perpendicular polarization arising from its spin cycloid, it fails to predict polarization for a proper screw. 
In NiI$_2$, the KNB scenario is partially satisfied because the canted screw along $\langle 1\bar{1}0\rangle$ includes a cycloidal component. However, when the proper screw states propagate along $\langle 110 \rangle$ directions, the magnetic point group ($2.1’$) enforces a polarization parallel to the propagation vector [Fig. \ref{1}(c4)], and the KNB model fails. 
Similarly, the KM model, $\mathbf{P}_{\rm KM} \propto \mathbf{S}_i \times \mathbf{S}_j$ \cite{zhang2012ordering}, which applies to proper screws, also proves only partially correct for the canted screw state of  NiI$_2$.
Hence, a comprehensive megnetoelectric model, capable of describing all spin–polarization configurations in NiI$_2$, is still highly desired.

To obtain an accurate magnetoelectric model for NiI$_2$, the generalized spin-current (GSC) method offers a promising avenue \cite{gKNB}. 
This method incorporates the complete interactions between spins and electric dipole via $\mathbf{P}_{\rm GSC} = \sum_{\alpha\beta} \mathbf{P}_{ij}^{\alpha\beta} S_{i\alpha}S_{j\beta}$, 
where the coefficients $\mathbf{P}_{ij}^{\alpha\beta}$ can be determined by fitting to \textit{ab initio} results. 
Because of its generality, the aforementioned models become special cases of the GSC approach, once the crystal symmetry is taken into account. 
Indeed, the GSC method has been applied to a range of materials and has successfully described their magnetoelectricity, 
including CuFeO$_2$ \cite{zhu2024mechanism}, MnI$_2$ \cite{gKNB}, AgCrO$_2$ \cite{gKNB}, and VI$_2$ \cite{VI2,yu2024interlayer},  by considering contributions from only the first nearest neighbors (NN).
For NiI$_2$, the GSC method has also been implemented, again relying on first nearest-neighbor (1NN) contributions, yet resulting in notable numerical discrepancies \cite{NiI2Nature,yu2025microscopic} (see also our results below). 
This suggests that further-neighbor interactions may be significant, 
especially since that the third NN (3NN) exchange $J_3$ can be comparable to the first NN exchange $J_1$. 
If additional exchange pathways are indeed relevant, questions naturally arise regarding their microscopic origins, 
how SOC operates over such distances, and whether this can be exploited to engineer stronger magnetoelectricity.

In this Letter, the GSC method, combined with density functional theory (DFT) calculations and a tight-binding (TB) model, is applied to investigate the magnetoelectric behavior of NiI$_2$ and its electronic origins. Besides the 1NN coupling, the 3NN interaction is found to be even stronger; Including both ensures that the model accurately reproduces the polarization for any spin configurations, not just the magnetic ground state. Interestingly, this phenomenon also appears in NiBr$_2$ but not in VI$_2$ or MnI$_2$. A TB analysis confirms a first-order dependence of polarization on the SOC strength and attributes the enhanced third NN coupling to stronger $p$--$d$ hopping involving $e_g$ orbitals. Moreover, changes in the spin texture under applied magnetic fields are clarified. Leveraging the present improved model, a complex pattern of polar vortex lattice is predicted.

\textcolor{blue}{Magnetoelectric model.} We start with applying the GSC method to NiI$_2$. The presence of inversion centers at the midpoints of 1NN to 3NN enables the simplification of the GSC model as follows,
\begin{equation}
\mathbf{P}_{\rm GSC} = \sum_{\langle ij \rangle _n} \mathcal{M}_{ij}^{n}  (\mathbf{S}_i \times \mathbf{S}_j)
\label{eq1}
\end{equation}
where $\mathcal{M}_{ij}^{n}$ is a $3\times 3$ matrix connecting the $n$th NN of spins $\mathbf{S}_i$ and $\mathbf{S}_j$. The elements of $\mathcal{M}$ can be determined by the four-state mapping method and DFT calculations \cite{xiang2013unified,sm}. With the $D_{3d}$ symmetry of NiI$_2$, the relevant $\mathcal{M}$ matrices for pairs along the $x$-axis (Fig. \ref{1}) are found to be (in units of $10^{-5}$\,\textit{e}\AA):
\[
    \mathcal{M}^{1} = \left[\begin{matrix}
11 & 0 & 0 \\
0 & 109 & 139 \\
0 & -4 & 5 
\end{matrix}\right],~~
\mathcal{M}^{3} = \left[\begin{matrix}
-1 & 0 & 0 \\
0 & 16 & 164 \\
0 & -12 & -10 
\end{matrix}\right].
\]
As one can see, the dominant elements include $M^1_{yy}$ and $M^1_{yz}$ for the 1NN matrix $ \mathcal{M}^{1}$. 
For instance, a positive $M^1_{yy}$ implies that, if $\mathbf{S}_i \parallel z$ and $\mathbf{S}_j \parallel x$, 
then an electric dipole forms along $y$. 
Likewise, a positive $M^1_{yz}$ indicates that $\mathbf{S}_i \parallel x$ and $\mathbf{S}_j \parallel y$ induce a $y$-directed dipole. 
Notably, the KNB model corresponds to an antisymmetric matrix with only nonzero entries of $M^{\rm KNB}_{yz}=-M^{\rm KNB}_{zy}$, 
which can be determined as $(M^1_{yz}-M^1_{zy})/2=72$. 
Hence, the KNB model actually omits, at least, the major diagonal contribution from $M^1_{yy}$.
Interestingly, though the $ \mathcal{M}^{2}$ is found to be negligible, the 3NN  matrix $ \mathcal{M}^{3}$ is significant, featuring the largest element $M^3_{yz}=164$ among all the $\mathcal{M}^{n}$ components.
To confirm that $M^3_{yz}$ indeed exceeds $M^1_{yy}$ and $M^1_{yz}$, tests are performed over a variety of pseudopotentials and Hubbard $U$ values, all of which consistently show $M^3_{yz} > M^1_{yy}, M^1_{yz}$ [see Table S2 in Supplementary Materials (SM) \cite{sm}].

\begin{figure}[t]
\includegraphics[width=8.5cm]{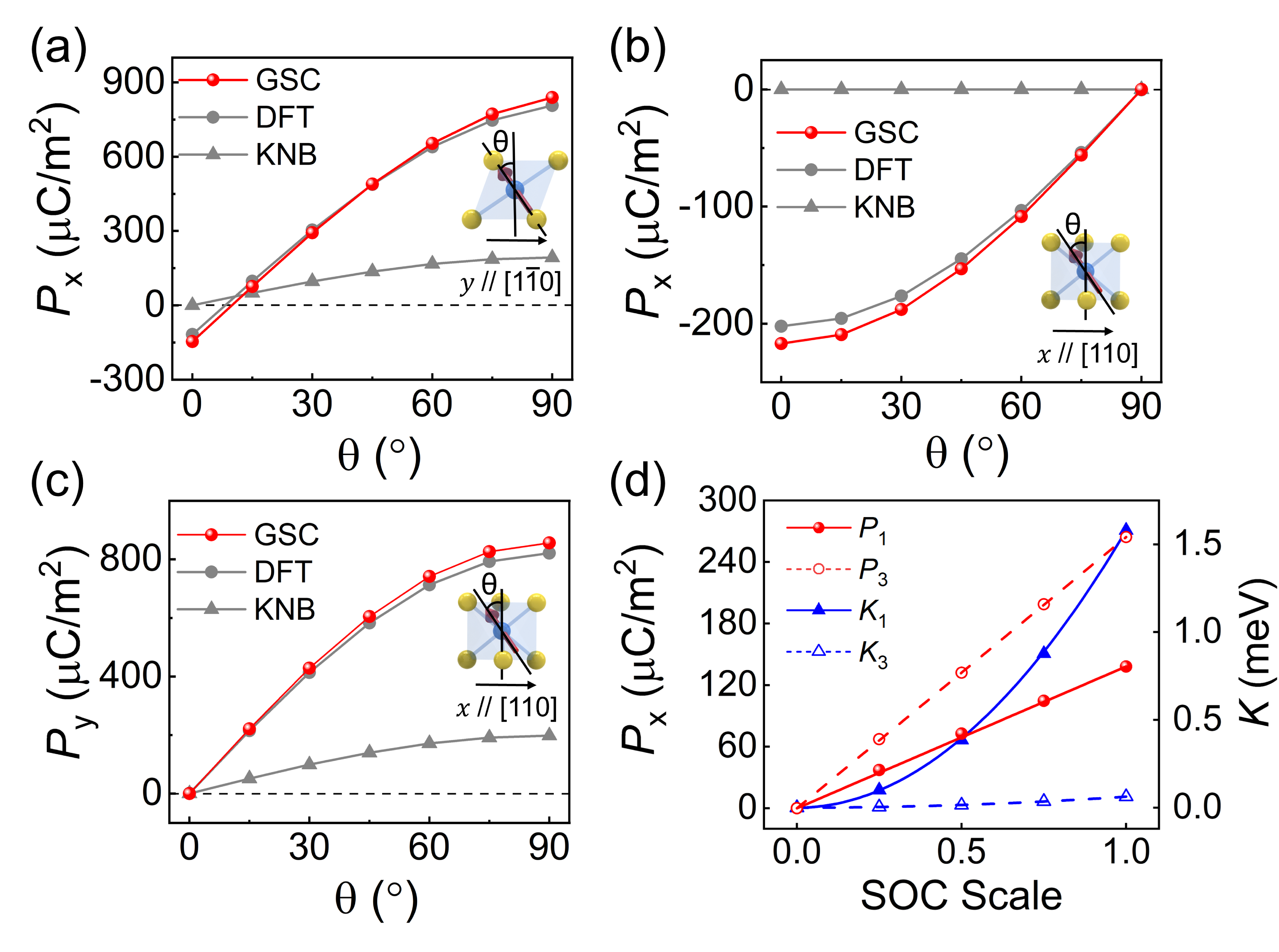}
\caption{Spin-induced polarization and its SOC dependence. (a) The $x$-component of polarization for a screw state propagating along $y$, plotted against the canting angle $\theta$. (b) and (c) show the $x$- and 
$y$-components of polarization, respectively, for a screw state propagating along $x$. 
The $P_z$ component is negligible and thus not shown here. In (a)--(c), data from the GSC model, 
DFT, and the KNB model are compared; the arrows indicate the propagation 
directions. (d) The polarization of a canted screw [along $y$, see Fig. \ref{1}(c1)] contributed by 
1NN and 3NN spin pairs, as well as 1NN and 3NN Kitaev 
interactions, as a function of the SOC scale.} 
\label{2}
\end{figure}

The accuracy of different models is now assessed by examining a screw state propagating along the $y$ direction (one of the $\langle 1\bar{1}0\rangle$ directions). 
As the canting angle of spin rotation plane $\theta$ evolves from $0^\circ$ to $90^\circ$, 
the state transforms from a proper screw, through canted screw states, and ultimately approaches an in-plane cycloid. 
Figure \ref{2}(a) shows that the GSC model including $\mathcal{M}^{1,3}$ produces a noticeable $x$-component of polarization, which is perpendicular to the propagation direction. 
For the proper screw at $\theta=0^\circ$, 
the GSC model predicts $P_x=-145.5$ $\mu$C/m$^2$. 
As $\theta$ increases from $0^\circ$ to $90^\circ$, $P_x$ reverses sign and continues to increase. 
In particular, at the characteristic angle $\theta=35.3^\circ$ determined by the Kitaev interaction, which defines the  magnetic ground state of bulk NiI$_{2}$\cite{NiI2Xu},  
the present GSC model yields $P_x=424.2$ $\mu$C/m$^2$. 
Moreover, the GSC model  shows close agreement with DFT for all $\theta$, 
indicating that the former is highly accurate. 
In contrast, the KNB model deviate significantly from the DFT results [Fig.\ref{2}(a)].



Furthermore, we also investigate a screw state propagating along the $x$ dirtection (one of the $\langle 110\rangle$ directions).
As shown in Figs.~\ref{2}(b) and (c), this 
$\langle 110\rangle$-propagated screw state exhibits both parallel and 
perpendicular polarizations that agree well with DFT. Specifically, the parallel 
component \(P_{x}\) reaches its maximum value of 
\(-217.1\,\mu\mathrm{C}/\mathrm{m}^2\) at \(\theta = 0^\circ\) and vanishes at 
\(\theta = 90^\circ\), whereas the KNB model predicts zero polarization for all 
\(\theta\). For the perpendicular component \(P_{y}\), the GSC model yields zero 
polarization at \(\theta = 0^\circ\) and attains a maximum of 
\(856.6\,\mu\mathrm{C}/\mathrm{m}^2\) at \(\theta = 90^\circ\). In contrast, the 
KNB model gives much smaller values than DFT does. These findings demonstrate that the 
induced polarization shifts continuously from being purely parallel to purely 
perpendicular to the propagation direction. Such transition is accurately captured by 
the GSC model, but not by either the KNB or KM models. Consequently, the GSC model 
with contributions from 1NN and 3NN, provides a comprehensive and precise  description of the multiferroicity in NiI$_2$.

We now examine how polarization depends on SOC by tuning 
SOC strength in DFT. The \(\mathcal{M}^{1,3}\) components are then updated to 
compute the 1NN and 3NN polarizations. Our calculations adopt 
the spin state shown in Fig.~\ref{1}(c1), which propagates along 
\(\langle 1\overline{1}0\rangle\). Its specific anisotropy arises from the interplay of the Kitaev 
interaction and single-ion anisotropy \cite{li2024effects}. As shown in 
Fig.~\ref{2}(d), \(\mathcal{M}^{3}\) contributes to polarization about 50\% more than 
\(\mathcal{M}^{1}\) does, further indicating the importnace of the 3NN. Both polarization contributions from 1NN and 3NN scale linearly with the SOC strength, indicating that 
SOC is crucial and extends beyond the nearest neighbors. Interestingly, despite a 
large \(J_{3}/J_{1}\) ratio for the Heisenberg term \cite{NiI2Xu}, the 3NN 
Kitaev interaction, which also arises from SOC, is found to be negligible 
[Fig.~\ref{2}(d)]. 

To understand the long-range impact of SOC, we investigate similar 
compounds, including NiBr$_{2}$ \cite{yu2025microscopic}, VI$_{2}$ \cite{VI2}, 
and MnI$_{2}$ \cite{gKNB}, which all exhibit polarization induced by noncollinear spin textures. The dominant $\mathcal{M}$ elements are summarized in 
Table~\ref{t1}. It shows that NiBr$_2$, possessing the same 3$d^8$ occupation as 
NiI$_2$, also exhibits $M^3_{yz} > M^1_{yy}, M^1_{yz}$, though these values are about an 
order smaller than those of NiI$_2$ due to weaker SOC of Br. In contrast, for MnI$_{2}$ 
(3$d^{5}$ high spin) and VI$_{2}$ (3$d^{3}$ high spin), $M^1_{yy}$ and $M^1_{yz}$ 
are already small, and $M^3_{yz}$ is  even smaller of an order. These findings suggest that 
$d$-orbital occupation plays a key role in 
producing the large SOC-induced $\mathcal{M}^{3}$ contribution.

\begin{table}[t]\centering
\caption{Dominant $\mathcal{M}$ matrix elements in various compounds. The 1NN and 3NN pairs are both aligned along $x$ direction [see Fig. \ref{1}(c2)].  Unit is in $10^{-5}$ $e$\AA.}
\renewcommand\arraystretch{1.22}
 \begin{tabular*}{8.5cm}{@{\quad}c@{\quad}r@{\quad}r@{\quad}r@{\quad} | @{\quad}c@{\quad}r@{\quad}r@{\quad}r}
\hline\hline
 & $M^1_{yy}$ & $M^1_{yz}$ & $M^3_{yz}$ & & $M^1_{yy}$ & $M^1_{yz}$ & $M^3_{yz}$\\
\hline
NiI$_2$  & 109 & 139 & 164 & MnI$_2$  & -24 & -26 & 4   \\
NiBr$_2$ &  9  & 16  & 23  & VI$_2$   & -19 & 13  & 2   \\
\hline
\hline
\end{tabular*}
\label{t1}  
\end{table}





\begin{figure}[tbh]
  \centering
  \includegraphics[width=\linewidth]{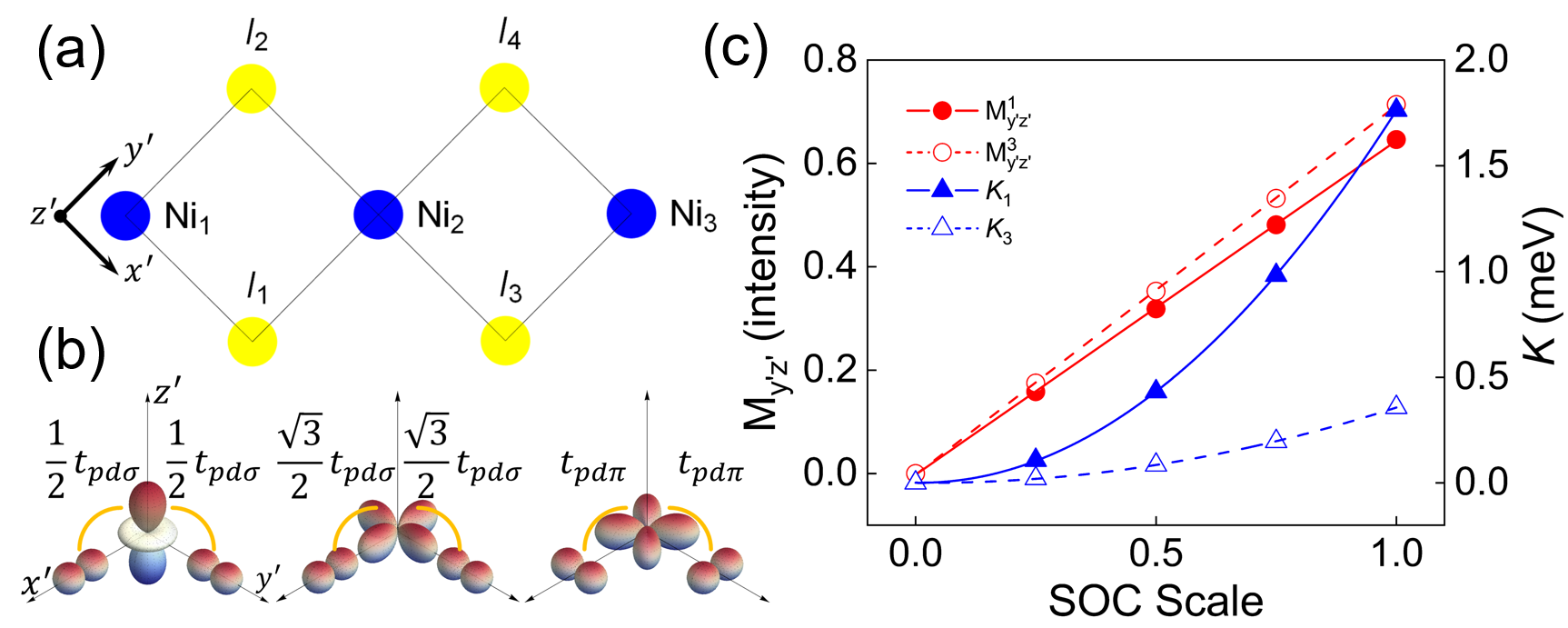}%
  \caption{The setups and results of TB model. Panel (a) schematizes the cluster used in the TB model, containing three Ni ions  and four ligand I anions . (b) Three types of hoppings used in the model.  (c) Variation of $M_{y'z'}$, as well as Kitaev interaction, for 1NN and 3NN  as a function of SOC scale. Note that only $e_g$ orbitals are considered for NiI$_2$ in the hole representation. }
  \label{3}
\end{figure}

\textcolor{blue}{Tight-binding interpretation.} To clarify the electronic origins of the large $\mathcal{M}^{3}$, we construct a tight-binding (TB) model.
As shown in Fig. \ref{3}(a), a cluster containing three Ni ions and four ligand I anions are considered.  The Ni-$d$ and I-$p$ orbitals are included, which appear near the Fermi level as determined by DFT calculations. We adopt a similar approach as in Refs. \cite{KNB,gKNB}, and express the Hamiltonian $\hat{\mathcal{H}}$ in three parts,
\begin{equation}
\hat{\mathcal{H}}=\hat{H}_{\rm Ni}+\hat{H}_{\rm I}+\hat{H}_t
\end{equation}
where the first part can be further expanded as $\hat{H}_{\rm Ni} = \hat{H}_{\rm Ni}^{\rm on}+\hat{H}_{\rm Ni}^{\rm SOC}+\hat{H}_{\rm Ni}^{U}$, which are onsite, SOC and Hubbard $U$ terms of Ni, respectively. The second part can similarly expanded as $\hat{H}_{\rm I} = \hat{H}_{\rm I}^{\rm on}+\hat{H}_{\rm I}^{\rm SOC}$ for ligand I.  The third part represents the hopping process,
\begin{equation}
\hat{H}_t=\sum_{i=1}^3\sum_{j=1}^4\sum_{\alpha\beta}\sum_{\sigma=\uparrow,\downarrow}\left(t_{i\alpha j\beta}\hat{p}_{b\beta\sigma}^{\dagger}\hat{d}_{a\alpha\sigma}+\mathrm{h.c.},\right)
\end{equation}
where the sums over $i,j$ run over Ni and I, respectively, while the indices ($\alpha,\beta$) and $\sigma$ represent orbital and spin. The hopping term $t_{i\alpha j\beta}$ is constructed with the Slater-Koster integrals $t_{pd\sigma}$ and $t_{pd\pi}$, see Fig. \ref{3}(b) and Table \ref{t3} in End Matter. Note that details of such TB model can be found in End Matter.


\begin{table}[b]\centering
\caption{Orbital-resolved contributions to the dominant \(M_{y'z'}\) elements.
This table shows the intensities of each component \(\langle \hat{d}_\alpha \mid e\vec{r'} \mid \hat{p}_\beta \rangle\) in the TB coordinates \(\{x', y', z'\}\) for 1NN and 3NN spin pairs. Here \(e\) is the elementary charge and $\vec{r'}$ is the corresponding position. The \(t_{2g}\) orbitals correspond to the case of VI\(_2\), and the \(e_g\) orbitals to the hole representation of NiI\(_2\).}
\begin{tabular}{l|rrrr}
\hline\hline
\quad\quad\quad\quad$M_{y'z'}$ \quad &\multicolumn{2}{c}{1NN}      & \multicolumn{2}{|c}{3NN} \\ 
\hline
\quad$\sum_{\beta}\bra{\hat{d}_{x'^2-y'^2}}e\vec{r'}\ket{\hat{p}_{\beta}}$ &\quad\quad-1.09\quad\quad &\multicolumn{1}{c|}{\multirow{2}{*}{0.65 }}&\quad \quad0.30\quad \quad &\multicolumn{1}{c}{\multirow{2}{*}{0.71 }}\\ 
\quad$\sum_{\beta}\bra{\hat{d}_{z'^2}}e\vec{r'}\ket{\hat{p}_{\beta}}$ & \quad\quad1.74\quad\quad &\multicolumn{1}{c|}{} &\quad\quad0.41\quad\quad &\multicolumn{1}{c}{}\\ 
\hline
\quad$\sum_{\beta}\bra{\hat{d}_{x'y'}}e\vec{r'}\ket{\hat{p}_{\beta}}$ & \quad\quad0.61\quad\quad &\multicolumn{1}{c|}{\multirow{3}{*}{0.60 }}& \quad\quad-0.13\quad\quad &\multicolumn{1}{c}{\multirow{3}{*}{-0.12 }} \\
\quad$\sum_{\beta}\bra{\hat{d}_{y'z'}}e\vec{r'}\ket{\hat{p}_{\beta}}$ &\quad\quad -0.01\quad\quad &\multicolumn{1}{c|}{} & \quad\quad0.00\quad \quad &\multicolumn{1}{c}{}\\ 
\quad$\sum_{\beta}\bra{\hat{d}_{z'x'}}e\vec{r'}\ket{\hat{p}_{\beta}}$ & \quad\quad0.00\quad \quad &\multicolumn{1}{c|}{} & \quad\quad0.01\quad \quad &\multicolumn{1}{c}{}\\
\hline\hline
\end{tabular}
\label{t2}
\end{table}

\begin{figure}[ht]
  \centering
  \includegraphics[width=\linewidth]{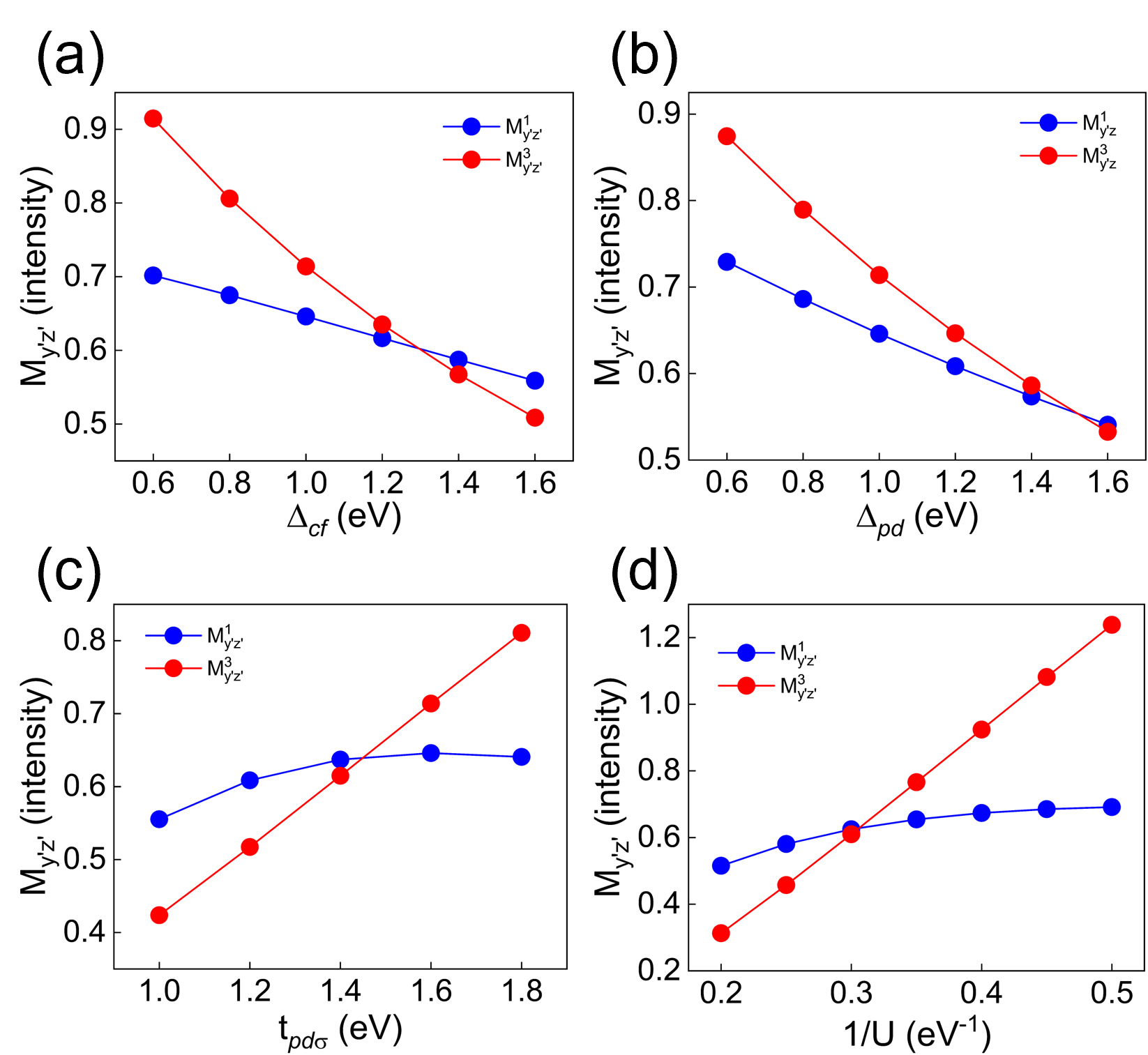}%
  \caption{Dependence of the $M^1_{y'z'}$ and $M^3_{y'z'}$ on tight binding parameters: (a) charge transfer energy $\Delta_{pd}$, (b) crystal field spiliting $\Delta_{\rm CF}$, (c) hopping integral  $t_{pd\sigma}$ and (d) Hubbard $U$ term.}
  \label{v2}
\end{figure}

For each spin configuration \(K\), we diagonalize \(\mathcal{H}\) to obtain the 
ground state \(\ket{G_K}\). Summing \(\bra{G_K}e\vec{r'}\ket{G_K}\) over all 
electrons yields the electronic dipole, from which the 
\(\mathcal{M}_{\mathrm{TB}}^n\) coefficients are determined via the four-state 
mapping method. Within this TB framework, the ratio 
\(\mu = M_{y'z'}^3 / M_{y'z'}^1\) is found to be 1.09, while 
\(\nu = K_3 / K_1\) is merely 0.20. In addition, the linear dependence of \(M\) and the quadratic dependence of Kitaev interaction \(K\) on spin-orbit coupling  are both well reproduced [Fig.~\ref{3}(c)], in 
excellent agreement with DFT. 
Furthermore, the ratios $\mu$ and $\nu$  are found robust
against variations in SOC scale $\lambda$, charge-transfer energy $\Delta_{pd}$, crystal
field $\Delta_{cf}$, hopping integrals and Hubbard $U$ [Fig.~\ref{v2}].



To gain a deeper understanding, we examine the orbital-resolved contributions to 
the \(M\) parameters. In our model, NiI\(_2\) is treated in the hole 
representation for the high-spin \(d^8\)  state, effectively placing two holes in 
the \(e_g\) orbitals. As shown in Table~\ref{t2}, the dominant contributions to 
\(\bra{\hat{d}_{\alpha}} e \vec{r'} \ket{\hat{p}_{\beta}}\) indeed originate 
from these \(e_g\) orbitals and remain significant up to 3NN. By contrast, in a system with three electrons (or holes) in the 
\(t_{2g}\) orbitals (as in VI\(_2\)), \(M_{y'z'}\) is already small at 1NN and becomes negligible at 3NN, indicating a much 
weaker long-range effect.
These differences arise from the edge-sharing octahedral geometry, where 
\(e_g\)--\(p\) hopping  is substantially larger than the 
 \(t_{2g}\)--\(p\) hopping ($t_{pd\sigma}>t_{pd\pi}$). As a result, NiI\(_2\) benefits 
from stronger long-range SOC effects, giving rise to a sizable \(M^3_{y'z'}\). 
Conversely, the Kitaev interaction, which is more sensitive to the \(t_{2g}\)--\(p\) 
hopping ($pd\pi$), quickly diminishes over distance. Moreover, Table \ref{t2} suggests that 
CoI\(_2\), possessing a high-spin \(d^7\) configuration, may exhibit robust and even stronger 
\(M\) values for both 1NN and 3NN.

 \begin{figure}[t]
\includegraphics[width=8.5cm]{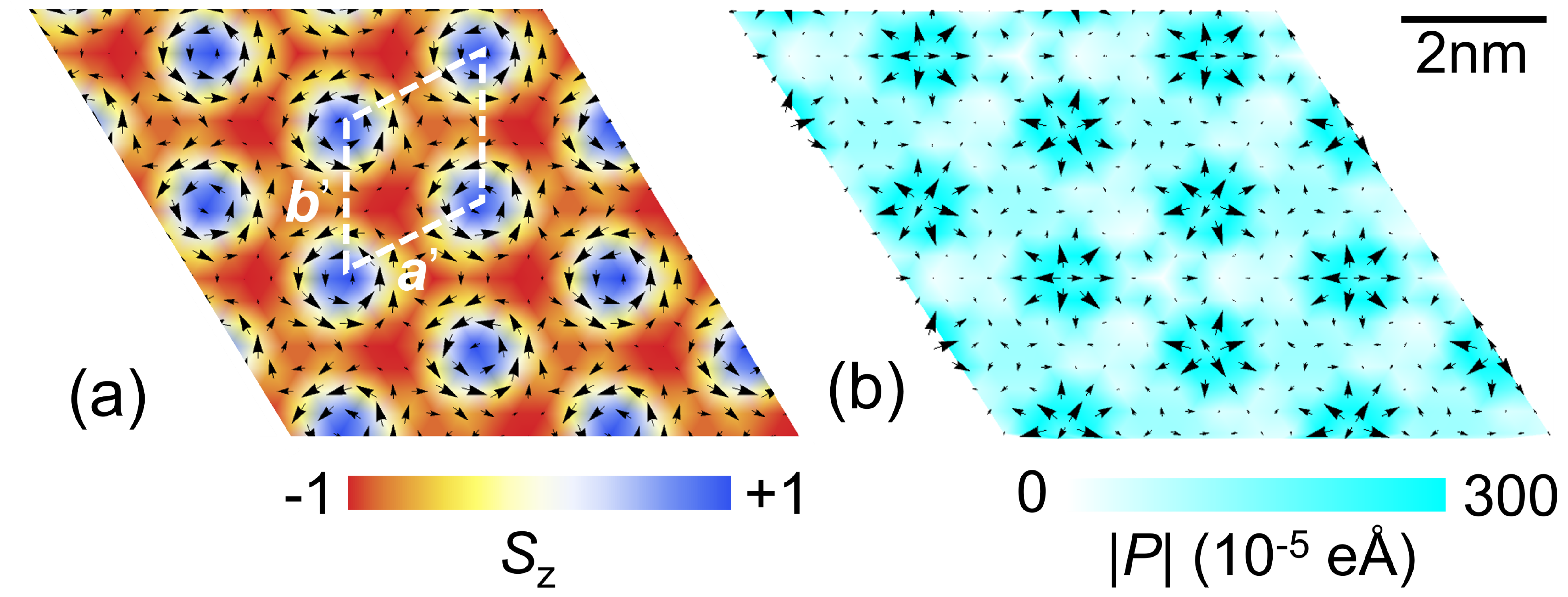}
\caption{ Magnetic skyrmion lattice and polar vortex pattern in monolayer NiI$_2$.  
Panel (a) the Bloch-type skyrmion lattice under a 30~T field, where the black arrows denote in-plane spins and the warm-cold color map depicts the out-of-plane spin component. The vectors \(\mathbf{a}'\) and \(\mathbf{b}'\) mark the skyrmion lattice unit cell. (b) The corresponding spin-induced polar vortex pattern, with black arrows showing the in-plane polarization and the color scale indicating the overall polarization magnitude (which is predominantly in-plane).  
} 
\label{4}
\end{figure}

\textcolor{blue}{Polar vortex lattice.}
Lastly, we apply the GSC model to explore polar textures in monolayer NiI$_2$. 
Starting from the monolayer ground state, \(\langle 110\rangle\)-propagating canted cycloid 
[Fig.~\ref{1}(c2)], we apply an out-of-plane magnetic field $B$ in Monte Carlo 
simulations (see SM for the spin Hamiltonian \cite{sm}). For $B \approx 30$ T,
a Bloch-type skyrmion lattice emerges [Fig. \ref{4}(a)], where each skyrmion 
carries topological charge 1 and the lattice constant is about 5.8 times of the 
 crystal lattice constant \cite{note1}.
Applying the GSC model in Eq. \ref{eq1} to this spin texture predicts  polar vortex pattern
[Fig.~\ref{4}(b)]. The polarization lies mostly in-plane, pointing radially outward from 
each skyrmion core, with the maximum polarization occuring at the core. It decays significantly 
beyond the skyrmion region. These polar vortices form a triangular 
lattice mirroring the underlying skyrmion arrangement. Interestingly, if the chirality of the initial screw state switches (the induced polarization flips), the outward-pointing polar vortices become inward-pointing ones.

In summary, the magnetoelectric mechanism in type-II multiferroic NiI$_{2}$  is investigated by developing a first-principles-based GSC model. It is found that third-nearest-neighbor spin pairs contribute a polarization comparable to that of first-nearest neighbors, rooting in the stronger $pd\sigma$ hopping involving $e_g$ orbitals of Ni$^{2+}$. By including both 1NN and 3NN spin pairs, the model accurately reproduces polarization for a variety of spin patterns and captures diverse polarization textures, including a polar vortex lattice under moderate fields. These findings clarify the mechanism behind spin-induced polarization in NiI$_{2}$ and reveal new opportunities for exploring rich multiferroic phases in layered magnets.

\begin{acknowledgments}
The work at Fudan is supported by NSFC (Grant No. 12274082), National Key R\&D Program of China (Grant No. 2022YFA1402901), Shanghai Pilot Program for Basic Research-FuDan University 21TQ1400100 (Grant No. 23TQ017),  Shanghai Science and Technology Committee (Grant No. 23ZR1406600),  Innovation Program for Quantum Science and Technology (Grant No. 2024ZD0300102), and Xiaomi Young Talents Program. W.D. acknowledges the Basic Science Center Project of NSFC (grant no. 52388201), Innovation Program for Quantum Science and Technology (2023ZD0300500), the Beijing Advanced Innovation Center for Future Chip (ICFC), and the Beijing Advanced Innovation Center for Materials Genome Engineering. J.F. acknowledges the Scientific Research Project of Universities in Anhui Province - 2024AH040216.

\end{acknowledgments}

\onecolumngrid
\begin{center}
\textbf{End Matter}
\end{center}
\twocolumngrid

\textcolor{blue}{Setups of TB model.} The TB Hamiltonian is constructed as, \[\hat{\mathcal{H}}=\hat{H}_{\rm Ni}+\hat{H}_{\rm I}+\hat{H}_t.\]
The first term $\hat{H}_\mathrm{Ni} = \hat{H}_\mathrm{Ni}^{\rm on}+\hat{H}_\mathrm{Ni}^{\rm SOC}+\hat{H}_\mathrm{Ni}^{U}$, is used to describe Ni$^{2+}$, where $\hat{H}_\mathrm{Ni}^{\rm on}$ is on-site energy term of $d$ orbits:
\[
\hat{H}_\mathrm{Ni}^{\rm on}=\sum_{i=1}^3\sum_{\sigma=\uparrow,\downarrow}\left[\sum_{\alpha\in{t_{2g}}}E_{t_{2g}}\hat{d}_{i\alpha\sigma}^\dagger\hat{d}_{i\alpha\sigma}+\sum_{\alpha\in e_g}E_{e_g}\hat{d}_{i\alpha\sigma}^\dagger\hat{d}_{i\alpha\sigma}\right],
\]
$\hat{H}_\mathrm{Ni}^{\rm SOC}$ is the SOC effect of magnetic ions:
\[
 \hat{H}_\mathrm{Ni}^{\rm SOC}=\lambda_\mathrm{Ni}\sum_{i=1}^3(\mathbf{L}_{i}\cdot\mathbf{S}_{i}),
\]
and $\hat{H}_\mathrm{Ni}^{U}$ denotes the effective Zeeman energy, which originates from the local Coulomb repulsion and Hund coupling in the magnetically ordered phase: 
\[
 \hat{H}_\mathrm{Ni}^{U}=-\mathcal{U}\sum_{i=1}^3\sum_{\alpha\in{t_{2g}},e_{g}}\mathbf{m}_{i}\cdot\mathbf{S}_{i\alpha}.
\]
Similarly, the second term $\hat{H}_\mathrm{I} = \hat{H}_\mathrm{I}^{\rm on}+\hat{H}_\mathrm{I}^{\rm SOC}$, is used to describe ligand I$^-$, where $\hat{H}_\mathrm{I}^{\rm on}$ is on-site energy terms of $p$ orbits:
\[
\hat{H}_\mathrm{I}^{on}=E_{p}\sum_{j=1}^4\sum_{\beta\in{x,y,z}}\sum_{\sigma=\uparrow,\downarrow}\hat{p}_{j\beta\sigma}^{\dagger}\hat{p}_{j\beta\sigma},
\]
$\hat{H}_\mathrm{I}^{\rm SOC}$ is the SOC effect of ligand anions:
\[
\hat{H}_\mathrm{I}^{\rm SOC}=\lambda_\mathrm{I}\sum_{j=1}^4(\mathbf{L}_{j}\cdot\mathbf{S}_{j}).
\]
The hopping term $\hat{H}_t$ is introduced in the main text and the hopping integrals are shown in Table \ref{t3}. 

\begin{table}[h]\centering
\caption{Hopping integrals between magnetic ions and ligands in edge-sharing octahedra. The left side shows the case where the Ni–I bond is oriented along the $x'$-axis, while the right side illustrates the bond along the $y'$-axis.}
\begin{tabular}{c|cccc|cccc}
 \hline\hline
\quad &along $\hat{x}$ & \quad$p_x$ & $p_y$  \quad&$p_z$  & along $\hat{y}$& \quad$p_x$  & \quad$p_y$\quad& $p_z$  \\ \hline
\multicolumn{1}{c|}{\multirow{3}{*}{$t_{2g}$}} & $d_{xy}$    &\quad0  & $t_{pd\pi}$ &\quad0     &  $d_{xy}$    &\quad $t_{pd\pi}$ &  \quad0                             &\quad0     \\
\multicolumn{1}{c|}{} &$d_{yz}$    &\quad0         & \quad0    &\quad0     & $d_{yz}$    & \quad0    &  \quad0       & $t_{pd\pi}$ \\
\multicolumn{1}{c|}{} &$d_{zx}$    &\quad0  &\quad0     & $t_{pd\pi}$ & $d_{zx}$    &  \quad0   &       \quad0                        &   \quad0  \\ \hline
\quad \multirow{2}{*}{$e_g$} &$d_{x^2-y^2}$ &$\frac{\sqrt{3}t_{pd\sigma}}{2}$ & \quad0   &  \quad0  &$d_{x^2-y^2}$ & \quad0  & $\frac{\sqrt{3}t_{pd\sigma}}{2}$ &   \quad0  \\
&$d_{z^2}$    & $-\frac{t_{pd\sigma}}{2}$ &  \quad0   & \quad0    & $d_{z^2}$    &  \quad0   & $-\frac{t_{pd\sigma}}{2}$ & \quad0 \\  \hline \hline
\end{tabular}
\label{t3}
\end{table}

The following parameters are adopted for the TB simulations. The orbital energies $E_{t2g}= 0$ eV, $E_{eg}= 1$ eV, $E_p=-2$ eV are estimated from the DFT band structure. To facilitate the calculation, we switch to the hole representation: $E_{t2g}= 1$ eV, $E_{eg}= 0 $ eV, $E_p=2$ eV. The SOC coefficient $\lambda_\mathrm{Ni}=-0.02$ eV and $\lambda_\mathrm{I}=-0.27$ eV is extracted from DFT calculations. The Hund term $U=3$ eV, the hopping integrals $t_{pd\sigma}=1.6$ eV and $t_{pd\pi}=0.6$ eV are commonly used values \cite{gKNB,feng2016anisotropic}. The dependence of results on such parameters is shown in Fig. \ref{v2}.

\textcolor{blue}{Methods to obtain $\mathcal{M}$. }
As aformentioned in the introdcution, the local electric dipoe in GSC model can be calulated with
\[\mathbf{P}^{\rm GSC}_{ij} = \sum_{\alpha\beta} \mathbf{P}_{ij}^{\alpha\beta} S_{i\alpha}S_{j\beta}.\]
Applying spatial inversion symmetry and time-reversal symmetry \cite{gKNB}, it can be simplified as, 
\[\mathbf{P}^{\rm GSC}_{ij} = \mathcal{M}_{ij}  (\mathbf{S}_i \times \mathbf{S}_j),\]
with
\[
\mathcal{M} = 
\begin{pmatrix}
(P^{yz})_x & (P^{zx})_x & (P^{xy})_x \\
(P^{yz})_y & (P^{zx})_y & (P^{xy})_y \\
(P^{yz})_z & (P^{zx})_z & (P^{xy})_z
\end{pmatrix}.
\]
where the subscripts $ij$ for $\mathcal{M}$ and $P$ are omited for simplicity.

We employ the four-state mapping method to obtain the $\mathbf{P}_{\rm GSC}$ and $\mathcal{M}$ coefficients. For instance, to extract $M_{y'z'}=(P^{x'y'})_{y'}$, we compute the ground states ($\ket{G_I}$, $\ket{G_{II}}$, $\ket{G_{III}}$, and $\ket{G_{IV}}$) for four magnetic configurations: 

(I). $S_1$ along $+x'$ and $S_3$ along $+y'$;

(II). $S_1$ along $+x'$ and $S_3$ along $-y'$;

(III). $S_1$ along $-x'$ and $S_3$ along $+y'$;

(IV). $S_1$ along $-x'$ and $S_3$ along $-y'$.\\
Here, $S_2$ is always aligned along $+z'$ being perpendicular to $S_1$ and $S_3$, with $S=1$. The $M_{y'z'}$ can then be exracted with  
\[
M_{y'z'} = \frac{e}{4S^2} \left( 
\begin{aligned}
&\sum_{I} \bra{G_I} \hat{y} \ket{G_I} - \sum_{II} \bra{G_{II}} \hat{y} \ket{G_{II}}- \\
& \sum_{III} \bra{G_{III}} \hat{y} \ket{G_{III}} + \sum_{IV} \bra{G_{IV}} \hat{y} \ket{G_{IV}}
\end{aligned}
\right),
\]
where the sums run over degenerate ground state(s).
The adopted cluster shown in Fig. \ref{3}(a) exhibits $D_{2h}$ symmetry, which requires the $\mathcal{M}_{\rm TB}$ matrix in the form of
\[
\mathcal{M}_{\rm TB}=
\begin{bmatrix}
0 & 0 & \quad -M_{y'z'} \\
0 & 0 & \quad M_{y'z'} \\
-M_{z'y'} & \quad M_{z'y'} & 0
\end{bmatrix}.
\]
The $\mathcal{M}^{1,3}$ matrices are determined as follows:
\begin{align*}
\mathcal{M}^{1}_{\rm TB} &=
\begin{bmatrix}
0 & 0 &  -0.65 \\
0 & 0 & \quad0.65 \\
0.00 & -0.00 & 0
\end{bmatrix} ,\\
\mathcal{M}^{3}_{\rm TB} &=
\begin{bmatrix}
0 & 0 &  -0.71  \\
0 & 0 & \quad0.71 \\
0.02 & -0.02 & 0
\end{bmatrix}.
\end{align*}
Note that (i) only the relative values are physically meaningful and (ii) the use of other reasonable parameters will not qualitatively change the results that $M_{y'z'}^3$ is comparable to $M_{y'z'}^1$, as shown in Fig. \ref{v2}.


\begin{thebibliography}{44}%
\makeatletter
\providecommand \@ifxundefined [1]{%
 \@ifx{#1\undefined}
}%
\providecommand \@ifnum [1]{%
 \ifnum #1\expandafter \@firstoftwo
 \else \expandafter \@secondoftwo
 \fi
}%
\providecommand \@ifx [1]{%
 \ifx #1\expandafter \@firstoftwo
 \else \expandafter \@secondoftwo
 \fi
}%
\providecommand \natexlab [1]{#1}%
\providecommand \enquote  [1]{``#1''}%
\providecommand \bibnamefont  [1]{#1}%
\providecommand \bibfnamefont [1]{#1}%
\providecommand \citenamefont [1]{#1}%
\providecommand \href@noop [0]{\@secondoftwo}%
\providecommand \href [0]{\begingroup \@sanitize@url \@href}%
\providecommand \@href[1]{\@@startlink{#1}\@@href}%
\providecommand \@@href[1]{\endgroup#1\@@endlink}%
\providecommand \@sanitize@url [0]{\catcode `\\12\catcode `\$12\catcode
  `\&12\catcode `\#12\catcode `\^12\catcode `\_12\catcode `\%12\relax}%
\providecommand \@@startlink[1]{}%
\providecommand \@@endlink[0]{}%
\providecommand \url  [0]{\begingroup\@sanitize@url \@url }%
\providecommand \@url [1]{\endgroup\@href {#1}{\urlprefix }}%
\providecommand \urlprefix  [0]{URL }%
\providecommand \Eprint [0]{\href }%
\providecommand \doibase [0]{https://doi.org/}%
\providecommand \selectlanguage [0]{\@gobble}%
\providecommand \bibinfo  [0]{\@secondoftwo}%
\providecommand \bibfield  [0]{\@secondoftwo}%
\providecommand \translation [1]{[#1]}%
\providecommand \BibitemOpen [0]{}%
\providecommand \bibitemStop [0]{}%
\providecommand \bibitemNoStop [0]{.\EOS\space}%
\providecommand \EOS [0]{\spacefactor3000\relax}%
\providecommand \BibitemShut  [1]{\csname bibitem#1\endcsname}%
\let\auto@bib@innerbib\@empty
\bibitem [{\citenamefont {Fiebig}\ \emph {et~al.}(2016)\citenamefont {Fiebig},
  \citenamefont {Lottermoser}, \citenamefont {Meier},\ and\ \citenamefont
  {Trassin}}]{MF2}%
  \BibitemOpen
  \bibfield  {author} {\bibinfo {author} {\bibfnamefont {M.}~\bibnamefont
  {Fiebig}}, \bibinfo {author} {\bibfnamefont {T.}~\bibnamefont {Lottermoser}},
  \bibinfo {author} {\bibfnamefont {D.}~\bibnamefont {Meier}},\ and\ \bibinfo
  {author} {\bibfnamefont {M.}~\bibnamefont {Trassin}},\ }\bibfield  {title}
  {\bibinfo {title} {The evolution of multiferroics},\ }\href
  {https://www.nature.com/articles/natrevmats201646} {\bibfield  {journal}
  {\bibinfo  {journal} {Nat. Rev. Mater.}\ }\textbf {\bibinfo {volume} {1}},\
  \bibinfo {pages} {1} (\bibinfo {year} {2016})}\BibitemShut {NoStop}%
\bibitem [{\citenamefont {Spaldin}\ and\ \citenamefont {Ramesh}(2019)}]{MF3}%
  \BibitemOpen
  \bibfield  {author} {\bibinfo {author} {\bibfnamefont {N.~A.}\ \bibnamefont
  {Spaldin}}\ and\ \bibinfo {author} {\bibfnamefont {R.}~\bibnamefont
  {Ramesh}},\ }\bibfield  {title} {\bibinfo {title} {Advances in
  magnetoelectric multiferroics},\ }\href
  {https://www.nature.com/articles/s41563-018-0275-2} {\bibfield  {journal}
  {\bibinfo  {journal} {Nat. Mater.}\ }\textbf {\bibinfo {volume} {18}},\
  \bibinfo {pages} {203} (\bibinfo {year} {2019})}\BibitemShut {NoStop}%
\bibitem [{\citenamefont {Cheong}\ and\ \citenamefont {Mostovoy}(2007)}]{MF4}%
  \BibitemOpen
  \bibfield  {author} {\bibinfo {author} {\bibfnamefont {S.-W.}\ \bibnamefont
  {Cheong}}\ and\ \bibinfo {author} {\bibfnamefont {M.}~\bibnamefont
  {Mostovoy}},\ }\bibfield  {title} {\bibinfo {title} {Multiferroics: a
  magnetic twist for ferroelectricity},\ }\href
  {https://www.nature.com/articles/nmat1804} {\bibfield  {journal} {\bibinfo
  {journal} {Nat. Mater.}\ }\textbf {\bibinfo {volume} {6}},\ \bibinfo {pages}
  {13} (\bibinfo {year} {2007})}\BibitemShut {NoStop}%
\bibitem [{\citenamefont {Xu}\ \emph {et~al.}(2024)\citenamefont {Xu},
  \citenamefont {Yu}, \citenamefont {Wang},\ and\ \citenamefont {Xiang}}]{XCS}%
  \BibitemOpen
  \bibfield  {author} {\bibinfo {author} {\bibfnamefont {C.}~\bibnamefont
  {Xu}}, \bibinfo {author} {\bibfnamefont {H.}~\bibnamefont {Yu}}, \bibinfo
  {author} {\bibfnamefont {J.}~\bibnamefont {Wang}},\ and\ \bibinfo {author}
  {\bibfnamefont {H.}~\bibnamefont {Xiang}},\ }\bibfield  {title} {\bibinfo
  {title} {First-principles approaches to magnetoelectric multiferroics},\
  }\href {https://doi.org/10.1146/annurev-conmatphys-032922-102353} {\bibfield
  {journal} {\bibinfo  {journal} {Annu. Rev. Condens. Matter Phys.}\ }\textbf
  {\bibinfo {volume} {15}},\ \bibinfo {pages} {85} (\bibinfo {year}
  {2024})}\BibitemShut {NoStop}%
\bibitem [{\citenamefont {Kuindersma}\ \emph {et~al.}(1981)\citenamefont
  {Kuindersma}, \citenamefont {Sanchez},\ and\ \citenamefont
  {Haas}}]{kuindersma1981magnetic}%
  \BibitemOpen
  \bibfield  {author} {\bibinfo {author} {\bibfnamefont {S.}~\bibnamefont
  {Kuindersma}}, \bibinfo {author} {\bibfnamefont {J.}~\bibnamefont
  {Sanchez}},\ and\ \bibinfo {author} {\bibfnamefont {C.}~\bibnamefont
  {Haas}},\ }\bibfield  {title} {\bibinfo {title} {Magnetic and structural
  investigations on $\mathrm{NiI}_{2}$ and $\mathrm{CoI}_{2}$},\ }\href
  {https://www.sciencedirect.com/science/article/abs/pii/0378436381901005}
  {\bibfield  {journal} {\bibinfo  {journal} {Physica B+ C}\ }\textbf {\bibinfo
  {volume} {111}},\ \bibinfo {pages} {231} (\bibinfo {year}
  {1981})}\BibitemShut {NoStop}%
\bibitem [{\citenamefont {Kurumaji}\ \emph {et~al.}(2013)\citenamefont
  {Kurumaji}, \citenamefont {Seki}, \citenamefont {Ishiwata}, \citenamefont
  {Murakawa}, \citenamefont {Kaneko},\ and\ \citenamefont
  {Tokura}}]{kurumaji2013magnetoelectric}%
  \BibitemOpen
  \bibfield  {author} {\bibinfo {author} {\bibfnamefont {T.}~\bibnamefont
  {Kurumaji}}, \bibinfo {author} {\bibfnamefont {S.}~\bibnamefont {Seki}},
  \bibinfo {author} {\bibfnamefont {S.}~\bibnamefont {Ishiwata}}, \bibinfo
  {author} {\bibfnamefont {H.}~\bibnamefont {Murakawa}}, \bibinfo {author}
  {\bibfnamefont {Y.}~\bibnamefont {Kaneko}},\ and\ \bibinfo {author}
  {\bibfnamefont {Y.}~\bibnamefont {Tokura}},\ }\bibfield  {title} {\bibinfo
  {title} {Magnetoelectric responses induced by domain rearrangement and spin
  structural change in triangular-lattice helimagnets $\mathrm{NiI}_{2}$ and
  $\mathrm{CoI}_{2}$},\ }\href
  {https://link.aps.org/doi/10.1103/PhysRevB.87.014429} {\bibfield  {journal}
  {\bibinfo  {journal} {Phys. Rev. B}\ }\textbf {\bibinfo {volume} {87}},\
  \bibinfo {pages} {014429} (\bibinfo {year} {2013})}\BibitemShut {NoStop}%
\bibitem [{\citenamefont {Li}\ \emph {et~al.}(2023)\citenamefont {Li},
  \citenamefont {Xu}, \citenamefont {Liu}, \citenamefont {Li}, \citenamefont
  {Bellaiche},\ and\ \citenamefont {Xiang}}]{NiI2Xu}%
  \BibitemOpen
  \bibfield  {author} {\bibinfo {author} {\bibfnamefont {X.}~\bibnamefont
  {Li}}, \bibinfo {author} {\bibfnamefont {C.}~\bibnamefont {Xu}}, \bibinfo
  {author} {\bibfnamefont {B.}~\bibnamefont {Liu}}, \bibinfo {author}
  {\bibfnamefont {X.}~\bibnamefont {Li}}, \bibinfo {author} {\bibfnamefont
  {L.}~\bibnamefont {Bellaiche}},\ and\ \bibinfo {author} {\bibfnamefont
  {H.}~\bibnamefont {Xiang}},\ }\bibfield  {title} {\bibinfo {title} {Realistic
  spin model for multiferroic $\mathrm{NiI}_{2}$},\ }\href
  {https://journals.aps.org/prl/abstract/10.1103/PhysRevLett.131.036701}
  {\bibfield  {journal} {\bibinfo  {journal} {Phys. Rev. Lett.}\ }\textbf
  {\bibinfo {volume} {131}},\ \bibinfo {pages} {036701} (\bibinfo {year}
  {2023})}\BibitemShut {NoStop}%
\bibitem [{\citenamefont {Miao}\ \emph {et~al.}(2023)\citenamefont {Miao},
  \citenamefont {Liu}, \citenamefont {Zhang}, \citenamefont {Wang},
  \citenamefont {Ji},\ and\ \citenamefont {Fu}}]{NiI2Fu}%
  \BibitemOpen
  \bibfield  {author} {\bibinfo {author} {\bibfnamefont {M.-P.}\ \bibnamefont
  {Miao}}, \bibinfo {author} {\bibfnamefont {N.}~\bibnamefont {Liu}}, \bibinfo
  {author} {\bibfnamefont {W.-H.}\ \bibnamefont {Zhang}}, \bibinfo {author}
  {\bibfnamefont {D.-B.}\ \bibnamefont {Wang}}, \bibinfo {author}
  {\bibfnamefont {W.}~\bibnamefont {Ji}},\ and\ \bibinfo {author}
  {\bibfnamefont {Y.-S.}\ \bibnamefont {Fu}},\ }\bibfield  {title} {\bibinfo
  {title} {Spin-resolved imaging of atomic-scale helimagnetism in monolayer
  $\mathrm{NiI}_{2}$},\ }\href {https://arxiv.org/abs/2309.16526} {\bibfield
  {journal} {\bibinfo  {journal} {arXiv:2309.16526}\ } (\bibinfo {year}
  {2023})}\BibitemShut {NoStop}%
\bibitem [{\citenamefont {Amini}\ \emph {et~al.}(2024)\citenamefont {Amini},
  \citenamefont {Fumega}, \citenamefont {Gonz{\'a}lez-Herrero}, \citenamefont
  {Va{\v{n}}o}, \citenamefont {Kezilebieke}, \citenamefont {Lado},\ and\
  \citenamefont {Liljeroth}}]{NiI2AM}%
  \BibitemOpen
  \bibfield  {author} {\bibinfo {author} {\bibfnamefont {M.}~\bibnamefont
  {Amini}}, \bibinfo {author} {\bibfnamefont {A.~O.}\ \bibnamefont {Fumega}},
  \bibinfo {author} {\bibfnamefont {H.}~\bibnamefont {Gonz{\'a}lez-Herrero}},
  \bibinfo {author} {\bibfnamefont {V.}~\bibnamefont {Va{\v{n}}o}}, \bibinfo
  {author} {\bibfnamefont {S.}~\bibnamefont {Kezilebieke}}, \bibinfo {author}
  {\bibfnamefont {J.~L.}\ \bibnamefont {Lado}},\ and\ \bibinfo {author}
  {\bibfnamefont {P.}~\bibnamefont {Liljeroth}},\ }\bibfield  {title} {\bibinfo
  {title} {Atomic-scale visualization of multiferroicity in monolayer
  $\mathrm{NiI}_{2}$},\ }\href {https://doi.org/10.1002/adma.202311342}
  {\bibfield  {journal} {\bibinfo  {journal} {Adv. Mater.}\ }\textbf {\bibinfo
  {volume} {36}},\ \bibinfo {pages} {2311342} (\bibinfo {year}
  {2024})}\BibitemShut {NoStop}%
\bibitem [{\citenamefont {Song}\ \emph {et~al.}(2022)\citenamefont {Song},
  \citenamefont {Occhialini}, \citenamefont {Erge{\c{c}}en}, \citenamefont
  {Ilyas}, \citenamefont {Amoroso}, \citenamefont {Barone}, \citenamefont
  {Kapeghian}, \citenamefont {Watanabe}, \citenamefont {Taniguchi},
  \citenamefont {Botana} \emph {et~al.}}]{NiI2Nature}%
  \BibitemOpen
  \bibfield  {author} {\bibinfo {author} {\bibfnamefont {Q.}~\bibnamefont
  {Song}}, \bibinfo {author} {\bibfnamefont {C.~A.}\ \bibnamefont
  {Occhialini}}, \bibinfo {author} {\bibfnamefont {E.}~\bibnamefont
  {Erge{\c{c}}en}}, \bibinfo {author} {\bibfnamefont {B.}~\bibnamefont
  {Ilyas}}, \bibinfo {author} {\bibfnamefont {D.}~\bibnamefont {Amoroso}},
  \bibinfo {author} {\bibfnamefont {P.}~\bibnamefont {Barone}}, \bibinfo
  {author} {\bibfnamefont {J.}~\bibnamefont {Kapeghian}}, \bibinfo {author}
  {\bibfnamefont {K.}~\bibnamefont {Watanabe}}, \bibinfo {author}
  {\bibfnamefont {T.}~\bibnamefont {Taniguchi}}, \bibinfo {author}
  {\bibfnamefont {A.~S.}\ \bibnamefont {Botana}}, \emph {et~al.},\ }\bibfield
  {title} {\bibinfo {title} {Evidence for a single-layer van der waals
  multiferroic},\ }\href {https://www.nature.com/articles/s41586-021-04337-x}
  {\bibfield  {journal} {\bibinfo  {journal} {Nature}\ }\textbf {\bibinfo
  {volume} {602}},\ \bibinfo {pages} {601} (\bibinfo {year}
  {2022})}\BibitemShut {NoStop}%
\bibitem [{\citenamefont {Ju}\ \emph {et~al.}(2021)\citenamefont {Ju},
  \citenamefont {Lee}, \citenamefont {Kim}, \citenamefont {Choi}, \citenamefont
  {Roh}, \citenamefont {Son}, \citenamefont {Park}, \citenamefont {Kim},
  \citenamefont {Jung}, \citenamefont {Kim} \emph {et~al.}}]{NiI2Nano}%
  \BibitemOpen
  \bibfield  {author} {\bibinfo {author} {\bibfnamefont {H.}~\bibnamefont
  {Ju}}, \bibinfo {author} {\bibfnamefont {Y.}~\bibnamefont {Lee}}, \bibinfo
  {author} {\bibfnamefont {K.-T.}\ \bibnamefont {Kim}}, \bibinfo {author}
  {\bibfnamefont {I.~H.}\ \bibnamefont {Choi}}, \bibinfo {author}
  {\bibfnamefont {C.~J.}\ \bibnamefont {Roh}}, \bibinfo {author} {\bibfnamefont
  {S.}~\bibnamefont {Son}}, \bibinfo {author} {\bibfnamefont {P.}~\bibnamefont
  {Park}}, \bibinfo {author} {\bibfnamefont {J.~H.}\ \bibnamefont {Kim}},
  \bibinfo {author} {\bibfnamefont {T.~S.}\ \bibnamefont {Jung}}, \bibinfo
  {author} {\bibfnamefont {J.~H.}\ \bibnamefont {Kim}}, \emph {et~al.},\
  }\bibfield  {title} {\bibinfo {title} {Possible persistence of multiferroic
  order down to bilayer limit of van der waals material $\mathrm{NiI}_{2}$},\
  }\href {https://pubs.acs.org/doi/abs/10.1021/acs.nanolett.1c01095} {\bibfield
   {journal} {\bibinfo  {journal} {Nano Lett.}\ }\textbf {\bibinfo {volume}
  {21}},\ \bibinfo {pages} {5126} (\bibinfo {year} {2021})}\BibitemShut
  {NoStop}%
\bibitem [{\citenamefont {Arima}(2007)}]{arima2007ferroelectricity}%
  \BibitemOpen
  \bibfield  {author} {\bibinfo {author} {\bibfnamefont {T.-h.}\ \bibnamefont
  {Arima}},\ }\bibfield  {title} {\bibinfo {title} {Ferroelectricity induced by
  proper-screw type magnetic order},\ }\href
  {https://journals.jps.jp/doi/ref/10.1143/jpsj.76.073702} {\bibfield
  {journal} {\bibinfo  {journal} {J. Phys. Soc. Jpn.}\ }\textbf {\bibinfo
  {volume} {76}},\ \bibinfo {pages} {073702} (\bibinfo {year}
  {2007})}\BibitemShut {NoStop}%
\bibitem [{\citenamefont {Jia}\ \emph {et~al.}(2006)\citenamefont {Jia},
  \citenamefont {Onoda}, \citenamefont {Nagaosa},\ and\ \citenamefont
  {Han}}]{jia2006bond}%
  \BibitemOpen
  \bibfield  {author} {\bibinfo {author} {\bibfnamefont {C.}~\bibnamefont
  {Jia}}, \bibinfo {author} {\bibfnamefont {S.}~\bibnamefont {Onoda}}, \bibinfo
  {author} {\bibfnamefont {N.}~\bibnamefont {Nagaosa}},\ and\ \bibinfo {author}
  {\bibfnamefont {J.~H.}\ \bibnamefont {Han}},\ }\bibfield  {title} {\bibinfo
  {title} {Bond electronic polarization induced by spin},\ }\href
  {https://journals.aps.org/prb/abstract/10.1103/PhysRevB.74.224444} {\bibfield
   {journal} {\bibinfo  {journal} {Phys. Rev. B}\ }\textbf {\bibinfo {volume}
  {74}},\ \bibinfo {pages} {224444} (\bibinfo {year} {2006})}\BibitemShut
  {NoStop}%
\bibitem [{\citenamefont {Zhu}\ \emph {et~al.}(2025)\citenamefont {Zhu},
  \citenamefont {Wang}, \citenamefont {Zhu}, \citenamefont {Zhu}, \citenamefont
  {Li}, \citenamefont {Zhao}, \citenamefont {Xu},\ and\ \citenamefont
  {Xiang}}]{zhu2024mechanism}%
  \BibitemOpen
  \bibfield  {author} {\bibinfo {author} {\bibfnamefont {W.}~\bibnamefont
  {Zhu}}, \bibinfo {author} {\bibfnamefont {P.}~\bibnamefont {Wang}}, \bibinfo
  {author} {\bibfnamefont {H.}~\bibnamefont {Zhu}}, \bibinfo {author}
  {\bibfnamefont {H.}~\bibnamefont {Zhu}}, \bibinfo {author} {\bibfnamefont
  {X.}~\bibnamefont {Li}}, \bibinfo {author} {\bibfnamefont {J.}~\bibnamefont
  {Zhao}}, \bibinfo {author} {\bibfnamefont {C.}~\bibnamefont {Xu}},\ and\
  \bibinfo {author} {\bibfnamefont {H.}~\bibnamefont {Xiang}},\ }\bibfield
  {title} {\bibinfo {title} {Mechanism of type-$\mathrm{II}$ multiferroicity in
  pure and $\mathrm{Al}$-doped $\mathrm{CuFeO}_{2}$},\ }\href
  {https://journals.aps.org/prl/abstract/10.1103/PhysRevLett.134.066801}
  {\bibfield  {journal} {\bibinfo  {journal} {Phys. Rev. Lett.}\ }\textbf
  {\bibinfo {volume} {134}},\ \bibinfo {pages} {066801} (\bibinfo {year}
  {2025})}\BibitemShut {NoStop}%
\bibitem [{\citenamefont {Feng}\ and\ \citenamefont
  {Xiang}(2016)}]{feng2016anisotropic}%
  \BibitemOpen
  \bibfield  {author} {\bibinfo {author} {\bibfnamefont {J.}~\bibnamefont
  {Feng}}\ and\ \bibinfo {author} {\bibfnamefont {H.}~\bibnamefont {Xiang}},\
  }\bibfield  {title} {\bibinfo {title} {Anisotropic symmetric exchange as a
  new mechanism for multiferroicity},\ }\href
  {https://journals.aps.org/prb/abstract/10.1103/PhysRevB.93.174416} {\bibfield
   {journal} {\bibinfo  {journal} {Phys. Rev. B}\ }\textbf {\bibinfo {volume}
  {93}},\ \bibinfo {pages} {174416} (\bibinfo {year} {2016})}\BibitemShut
  {NoStop}%
\bibitem [{\citenamefont {Fumega}\ and\ \citenamefont {Lado}(2022)}]{KNB2}%
  \BibitemOpen
  \bibfield  {author} {\bibinfo {author} {\bibfnamefont {A.~O.}\ \bibnamefont
  {Fumega}}\ and\ \bibinfo {author} {\bibfnamefont {J.}~\bibnamefont {Lado}},\
  }\bibfield  {title} {\bibinfo {title} {Microscopic origin of multiferroic
  order in monolayer $\mathrm{NiI}_{2}$},\ }\href
  {https://iopscience.iop.org/article/10.1088/2053-1583/ac4e9d/meta} {\bibfield
   {journal} {\bibinfo  {journal} {2D Mater.}\ }\textbf {\bibinfo {volume}
  {9}},\ \bibinfo {pages} {025010} (\bibinfo {year} {2022})}\BibitemShut
  {NoStop}%
\bibitem [{\citenamefont {Sergienko}\ \emph {et~al.}(2006)\citenamefont
  {Sergienko}, \citenamefont {{\c{S}}en},\ and\ \citenamefont
  {Dagotto}}]{sergienko2006ferroelectricity}%
  \BibitemOpen
  \bibfield  {author} {\bibinfo {author} {\bibfnamefont {I.~A.}\ \bibnamefont
  {Sergienko}}, \bibinfo {author} {\bibfnamefont {C.}~\bibnamefont
  {{\c{S}}en}},\ and\ \bibinfo {author} {\bibfnamefont {E.}~\bibnamefont
  {Dagotto}},\ }\bibfield  {title} {\bibinfo {title} {Ferroelectricity in the
  magnetic $\mathrm{E}$-phase of orthorhombic perovskites},\ }\href
  {https://journals.aps.org/prl/abstract/10.1103/PhysRevLett.97.227204}
  {\bibfield  {journal} {\bibinfo  {journal} {Phys. Rev. Lett.}\ }\textbf
  {\bibinfo {volume} {97}},\ \bibinfo {pages} {227204} (\bibinfo {year}
  {2006})}\BibitemShut {NoStop}%
\bibitem [{\citenamefont {Picozzi}\ \emph {et~al.}(2007)\citenamefont
  {Picozzi}, \citenamefont {Yamauchi}, \citenamefont {Sanyal}, \citenamefont
  {Sergienko},\ and\ \citenamefont {Dagotto}}]{picozzi2007dual}%
  \BibitemOpen
  \bibfield  {author} {\bibinfo {author} {\bibfnamefont {S.}~\bibnamefont
  {Picozzi}}, \bibinfo {author} {\bibfnamefont {K.}~\bibnamefont {Yamauchi}},
  \bibinfo {author} {\bibfnamefont {B.}~\bibnamefont {Sanyal}}, \bibinfo
  {author} {\bibfnamefont {I.~A.}\ \bibnamefont {Sergienko}},\ and\ \bibinfo
  {author} {\bibfnamefont {E.}~\bibnamefont {Dagotto}},\ }\bibfield  {title}
  {\bibinfo {title} {Dual nature of improper ferroelectricity in a
  magnetoelectric multiferroic},\ }\href
  {https://journals.aps.org/prl/abstract/10.1103/PhysRevLett.99.227201}
  {\bibfield  {journal} {\bibinfo  {journal} {Phys. Rev. Lett.}\ }\textbf
  {\bibinfo {volume} {99}},\ \bibinfo {pages} {227201} (\bibinfo {year}
  {2007})}\BibitemShut {NoStop}%
\bibitem [{\citenamefont {Wu}\ \emph {et~al.}(2024)\citenamefont {Wu},
  \citenamefont {Zeng}, \citenamefont {Lu}, \citenamefont {Han}, \citenamefont
  {Yang}, \citenamefont {Liu}, \citenamefont {Zhao}, \citenamefont {Qiao},
  \citenamefont {Ji}, \citenamefont {Che} \emph {et~al.}}]{KNB3}%
  \BibitemOpen
  \bibfield  {author} {\bibinfo {author} {\bibfnamefont {Y.}~\bibnamefont
  {Wu}}, \bibinfo {author} {\bibfnamefont {Z.}~\bibnamefont {Zeng}}, \bibinfo
  {author} {\bibfnamefont {H.}~\bibnamefont {Lu}}, \bibinfo {author}
  {\bibfnamefont {X.}~\bibnamefont {Han}}, \bibinfo {author} {\bibfnamefont
  {C.}~\bibnamefont {Yang}}, \bibinfo {author} {\bibfnamefont {N.}~\bibnamefont
  {Liu}}, \bibinfo {author} {\bibfnamefont {X.}~\bibnamefont {Zhao}}, \bibinfo
  {author} {\bibfnamefont {L.}~\bibnamefont {Qiao}}, \bibinfo {author}
  {\bibfnamefont {W.}~\bibnamefont {Ji}}, \bibinfo {author} {\bibfnamefont
  {R.}~\bibnamefont {Che}}, \emph {et~al.},\ }\bibfield  {title} {\bibinfo
  {title} {Coexistence of ferroelectricity and antiferroelectricity in
  2$\mathrm{D}$ van der waals multiferroic},\ }\href
  {https://www.nature.com/articles/s41467-024-53019-5} {\bibfield  {journal}
  {\bibinfo  {journal} {Nat. Commun.}\ }\textbf {\bibinfo {volume} {15}},\
  \bibinfo {pages} {8616} (\bibinfo {year} {2024})}\BibitemShut {NoStop}%
\bibitem [{\citenamefont {Ant{\~a}o}\ \emph {et~al.}(2024)\citenamefont
  {Ant{\~a}o}, \citenamefont {Lado},\ and\ \citenamefont
  {Fumega}}]{antao2024electric}%
  \BibitemOpen
  \bibfield  {author} {\bibinfo {author} {\bibfnamefont {T.~V.}\ \bibnamefont
  {Ant{\~a}o}}, \bibinfo {author} {\bibfnamefont {J.~L.}\ \bibnamefont
  {Lado}},\ and\ \bibinfo {author} {\bibfnamefont {A.~O.}\ \bibnamefont
  {Fumega}},\ }\bibfield  {title} {\bibinfo {title} {Electric field control of
  moir{\'e} skyrmion phases in twisted multiferroic $\mathrm{NiI}_{2}$
  bilayers},\ }\href {https://pubs.acs.org/doi/10.1021/acs.nanolett.4c04582}
  {\bibfield  {journal} {\bibinfo  {journal} {Nano Lett.}\ }\textbf {\bibinfo
  {volume} {24}},\ \bibinfo {pages} {15767} (\bibinfo {year}
  {2024})}\BibitemShut {NoStop}%
\bibitem [{\citenamefont {Katsura}\ \emph {et~al.}(2005)\citenamefont
  {Katsura}, \citenamefont {Nagaosa},\ and\ \citenamefont {Balatsky}}]{KNB}%
  \BibitemOpen
  \bibfield  {author} {\bibinfo {author} {\bibfnamefont {H.}~\bibnamefont
  {Katsura}}, \bibinfo {author} {\bibfnamefont {N.}~\bibnamefont {Nagaosa}},\
  and\ \bibinfo {author} {\bibfnamefont {A.~V.}\ \bibnamefont {Balatsky}},\
  }\bibfield  {title} {\bibinfo {title} {Spin current and magnetoelectric
  effect in noncollinear magnets},\ }\href
  {https://doi.org/10.1103/PhysRevLett.95.057205} {\bibfield  {journal}
  {\bibinfo  {journal} {Phys. Rev. Lett.}\ }\textbf {\bibinfo {volume} {95}},\
  \bibinfo {pages} {057205} (\bibinfo {year} {2005})}\BibitemShut {NoStop}%
\bibitem [{\citenamefont {Zhang}\ \emph {et~al.}(2012)\citenamefont {Zhang},
  \citenamefont {Singh}, \citenamefont {Guillou}, \citenamefont {Simon},
  \citenamefont {Breard}, \citenamefont {Caignaert},\ and\ \citenamefont
  {Hardy}}]{zhang2012ordering}%
  \BibitemOpen
  \bibfield  {author} {\bibinfo {author} {\bibfnamefont {Q.}~\bibnamefont
  {Zhang}}, \bibinfo {author} {\bibfnamefont {K.}~\bibnamefont {Singh}},
  \bibinfo {author} {\bibfnamefont {F.}~\bibnamefont {Guillou}}, \bibinfo
  {author} {\bibfnamefont {C.}~\bibnamefont {Simon}}, \bibinfo {author}
  {\bibfnamefont {Y.}~\bibnamefont {Breard}}, \bibinfo {author} {\bibfnamefont
  {V.}~\bibnamefont {Caignaert}},\ and\ \bibinfo {author} {\bibfnamefont
  {V.}~\bibnamefont {Hardy}},\ }\bibfield  {title} {\bibinfo {title} {Ordering
  process and ferroelectricity in a spinel derived from
  $\mathrm{FeV}_{2}\mathrm{O}_{4}$},\ }\href
  {https://journals.aps.org/prb/abstract/10.1103/PhysRevB.85.054405} {\bibfield
   {journal} {\bibinfo  {journal} {Phys. Rev. B}\ }\textbf {\bibinfo {volume}
  {85}},\ \bibinfo {pages} {054405} (\bibinfo {year} {2012})}\BibitemShut
  {NoStop}%
\bibitem [{\citenamefont {Xiang}\ \emph {et~al.}(2011)\citenamefont {Xiang},
  \citenamefont {Kan}, \citenamefont {Zhang}, \citenamefont {Whangbo},\ and\
  \citenamefont {Gong}}]{gKNB}%
  \BibitemOpen
  \bibfield  {author} {\bibinfo {author} {\bibfnamefont {H.~J.}\ \bibnamefont
  {Xiang}}, \bibinfo {author} {\bibfnamefont {E.~J.}\ \bibnamefont {Kan}},
  \bibinfo {author} {\bibfnamefont {Y.}~\bibnamefont {Zhang}}, \bibinfo
  {author} {\bibfnamefont {M.-H.}\ \bibnamefont {Whangbo}},\ and\ \bibinfo
  {author} {\bibfnamefont {X.~G.}\ \bibnamefont {Gong}},\ }\bibfield  {title}
  {\bibinfo {title} {General theory for the ferroelectric polarization induced
  by spin-spiral order},\ }\href
  {https://doi.org/10.1103/PhysRevLett.107.157202} {\bibfield  {journal}
  {\bibinfo  {journal} {Phys. Rev. Lett.}\ }\textbf {\bibinfo {volume} {107}},\
  \bibinfo {pages} {157202} (\bibinfo {year} {2011})}\BibitemShut {NoStop}%
\bibitem [{\citenamefont {Liu}\ \emph {et~al.}(2024{\natexlab{a}})\citenamefont
  {Liu}, \citenamefont {Ren},\ and\ \citenamefont {Picozzi}}]{VI2}%
  \BibitemOpen
  \bibfield  {author} {\bibinfo {author} {\bibfnamefont {C.}~\bibnamefont
  {Liu}}, \bibinfo {author} {\bibfnamefont {W.}~\bibnamefont {Ren}},\ and\
  \bibinfo {author} {\bibfnamefont {S.}~\bibnamefont {Picozzi}},\ }\bibfield
  {title} {\bibinfo {title} {Spin-chirality-driven multiferroicity in van der
  waals monolayers},\ }\href {https://doi.org/10.1103/PhysRevLett.132.086802}
  {\bibfield  {journal} {\bibinfo  {journal} {Phys. Rev. Lett.}\ }\textbf
  {\bibinfo {volume} {132}},\ \bibinfo {pages} {086802} (\bibinfo {year}
  {2024}{\natexlab{a}})}\BibitemShut {NoStop}%
\bibitem [{\citenamefont {Yu}\ \emph {et~al.}(2024)\citenamefont {Yu},
  \citenamefont {Xu}, \citenamefont {Dai}, \citenamefont {Sun}, \citenamefont
  {Huang},\ and\ \citenamefont {Wei}}]{yu2024interlayer}%
  \BibitemOpen
  \bibfield  {author} {\bibinfo {author} {\bibfnamefont {S.}~\bibnamefont
  {Yu}}, \bibinfo {author} {\bibfnamefont {Y.}~\bibnamefont {Xu}}, \bibinfo
  {author} {\bibfnamefont {Y.}~\bibnamefont {Dai}}, \bibinfo {author}
  {\bibfnamefont {D.}~\bibnamefont {Sun}}, \bibinfo {author} {\bibfnamefont
  {B.}~\bibnamefont {Huang}},\ and\ \bibinfo {author} {\bibfnamefont
  {W.}~\bibnamefont {Wei}},\ }\bibfield  {title} {\bibinfo {title} {Interlayer
  magnetoelectric coupling in van der waals structures},\ }\href
  {https://journals.aps.org/prb/abstract/10.1103/PhysRevB.109.L100402}
  {\bibfield  {journal} {\bibinfo  {journal} {Phys. Rev. B}\ }\textbf {\bibinfo
  {volume} {109}},\ \bibinfo {pages} {L100402} (\bibinfo {year}
  {2024})}\BibitemShut {NoStop}%
\bibitem [{\citenamefont {Yu}\ \emph {et~al.}(2025)\citenamefont {Yu},
  \citenamefont {Ni}, \citenamefont {Yao},\ and\ \citenamefont
  {Cao}}]{yu2025microscopic}%
  \BibitemOpen
  \bibfield  {author} {\bibinfo {author} {\bibfnamefont {H.-S.}\ \bibnamefont
  {Yu}}, \bibinfo {author} {\bibfnamefont {X.-S.}\ \bibnamefont {Ni}}, \bibinfo
  {author} {\bibfnamefont {D.-X.}\ \bibnamefont {Yao}},\ and\ \bibinfo {author}
  {\bibfnamefont {K.}~\bibnamefont {Cao}},\ }\bibfield  {title} {\bibinfo
  {title} {Microscopic origin of magnetoferroelectricity in monolayer
  $\mathrm{NiBr}_{2}$ and $\mathrm{NiI}_{2}$},\ }\href
  {https://arxiv.org/abs/2501.05025} {\bibfield  {journal} {\bibinfo  {journal}
  {arXiv:2501.05025}\ } (\bibinfo {year} {2025})}\BibitemShut {NoStop}%
\bibitem [{\citenamefont {Xiang}\ \emph {et~al.}(2013)\citenamefont {Xiang},
  \citenamefont {Wang}, \citenamefont {Whangbo},\ and\ \citenamefont
  {Gong}}]{xiang2013unified}%
  \BibitemOpen
  \bibfield  {author} {\bibinfo {author} {\bibfnamefont {H.}~\bibnamefont
  {Xiang}}, \bibinfo {author} {\bibfnamefont {P.}~\bibnamefont {Wang}},
  \bibinfo {author} {\bibfnamefont {M.-H.}\ \bibnamefont {Whangbo}},\ and\
  \bibinfo {author} {\bibfnamefont {X.}~\bibnamefont {Gong}},\ }\bibfield
  {title} {\bibinfo {title} {Unified model of ferroelectricity induced by spin
  order},\ }\href
  {https://journals.aps.org/prb/abstract/10.1103/PhysRevB.88.054404} {\bibfield
   {journal} {\bibinfo  {journal} {Phys. Rev. B}\ }\textbf {\bibinfo {volume}
  {88}},\ \bibinfo {pages} {054404} (\bibinfo {year} {2013})}\BibitemShut
  {NoStop}%
\bibitem [{sm()}]{sm}%
  \BibitemOpen
  \href@noop {} {}\bibinfo {note} {See Supplementary Materials for more details
  about calculation methods and further discussions, which includes
  Refs.\cite{NiI2Xu,VASP, PBE,NiI2Wei,U4.2,DFT-D3,BF,VI2,gKNB,
  PASP,Weiyi2,XCS2,NiI2Xiang,LXY,XCS,
  PTMC,CG,yu2025microscopic,NiI2Fu,NiI2Silvia,li2024effects}.}\BibitemShut
  {Stop}%
\bibitem [{\citenamefont {Li}\ \emph {et~al.}(2024)\citenamefont {Li},
  \citenamefont {Zhang}, \citenamefont {Chen}, \citenamefont {Xu},\ and\
  \citenamefont {Xiang}}]{li2024effects}%
  \BibitemOpen
  \bibfield  {author} {\bibinfo {author} {\bibfnamefont {L.}~\bibnamefont
  {Li}}, \bibinfo {author} {\bibfnamefont {B.}~\bibnamefont {Zhang}}, \bibinfo
  {author} {\bibfnamefont {Z.}~\bibnamefont {Chen}}, \bibinfo {author}
  {\bibfnamefont {C.}~\bibnamefont {Xu}},\ and\ \bibinfo {author}
  {\bibfnamefont {H.}~\bibnamefont {Xiang}},\ }\bibfield  {title} {\bibinfo
  {title} {Effects of kitaev interaction on magnetic order and anisotropy},\
  }\href {https://doi.org/10.1103/PhysRevB.110.214435} {\bibfield  {journal}
  {\bibinfo  {journal} {Phys. Rev. B}\ }\textbf {\bibinfo {volume} {110}},\
  \bibinfo {pages} {214435} (\bibinfo {year} {2024})}\BibitemShut {NoStop}%
\bibitem [{not()}]{note1}%
  \BibitemOpen
  \href@noop {} {}\bibinfo {note} {Such state is reministic of that in Ref.
  \cite{NiI2Silvia}, where biquadratic interaction is neglected and skyrmions
  exhibit topological charge 2.}\BibitemShut {Stop}%
\bibitem [{\citenamefont {Kresse}\ and\ \citenamefont
  {Furthm\"uller}(1996)}]{VASP}%
  \BibitemOpen
  \bibfield  {author} {\bibinfo {author} {\bibfnamefont {G.}~\bibnamefont
  {Kresse}}\ and\ \bibinfo {author} {\bibfnamefont {J.}~\bibnamefont
  {Furthm\"uller}},\ }\bibfield  {title} {\bibinfo {title} {Efficient iterative
  schemes for ab initio total-energy calculations using a plane-wave basis
  set},\ }\href {https://doi.org/10.1103/PhysRevB.54.11169} {\bibfield
  {journal} {\bibinfo  {journal} {Phys. Rev. B}\ }\textbf {\bibinfo {volume}
  {54}},\ \bibinfo {pages} {11169} (\bibinfo {year} {1996})}\BibitemShut
  {NoStop}%
\bibitem [{\citenamefont {Perdew}\ \emph {et~al.}(1996)\citenamefont {Perdew},
  \citenamefont {Burke},\ and\ \citenamefont {Ernzerhof}}]{PBE}%
  \BibitemOpen
  \bibfield  {author} {\bibinfo {author} {\bibfnamefont {J.~P.}\ \bibnamefont
  {Perdew}}, \bibinfo {author} {\bibfnamefont {K.}~\bibnamefont {Burke}},\ and\
  \bibinfo {author} {\bibfnamefont {M.}~\bibnamefont {Ernzerhof}},\ }\bibfield
  {title} {\bibinfo {title} {Generalized gradient approximation made simple},\
  }\href {https://doi.org/10.1103/PhysRevLett.77.3865} {\bibfield  {journal}
  {\bibinfo  {journal} {Phys. Rev. Lett.}\ }\textbf {\bibinfo {volume} {77}},\
  \bibinfo {pages} {3865} (\bibinfo {year} {1996})}\BibitemShut {NoStop}%
\bibitem [{\citenamefont {Liu}\ \emph {et~al.}(2024{\natexlab{b}})\citenamefont
  {Liu}, \citenamefont {Wang}, \citenamefont {Yan}, \citenamefont {Xu},
  \citenamefont {Hu}, \citenamefont {Zhang},\ and\ \citenamefont
  {Ji}}]{NiI2Wei}%
  \BibitemOpen
  \bibfield  {author} {\bibinfo {author} {\bibfnamefont {N.}~\bibnamefont
  {Liu}}, \bibinfo {author} {\bibfnamefont {C.}~\bibnamefont {Wang}}, \bibinfo
  {author} {\bibfnamefont {C.}~\bibnamefont {Yan}}, \bibinfo {author}
  {\bibfnamefont {C.}~\bibnamefont {Xu}}, \bibinfo {author} {\bibfnamefont
  {J.}~\bibnamefont {Hu}}, \bibinfo {author} {\bibfnamefont {Y.}~\bibnamefont
  {Zhang}},\ and\ \bibinfo {author} {\bibfnamefont {W.}~\bibnamefont {Ji}},\
  }\bibfield  {title} {\bibinfo {title} {Competing multiferroic phases in
  monolayer and few-layer $\mathrm{NiI}_{2}$},\ }\href
  {https://doi.org/10.1103/PhysRevB.109.195422} {\bibfield  {journal} {\bibinfo
   {journal} {Phys. Rev. B}\ }\textbf {\bibinfo {volume} {109}},\ \bibinfo
  {pages} {195422} (\bibinfo {year} {2024}{\natexlab{b}})}\BibitemShut
  {NoStop}%
\bibitem [{\citenamefont {Botana}\ and\ \citenamefont {Norman}(2019)}]{U4.2}%
  \BibitemOpen
  \bibfield  {author} {\bibinfo {author} {\bibfnamefont {A.~S.}\ \bibnamefont
  {Botana}}\ and\ \bibinfo {author} {\bibfnamefont {M.~R.}\ \bibnamefont
  {Norman}},\ }\bibfield  {title} {\bibinfo {title} {Electronic structure and
  magnetism of transition metal dihalides: Bulk to monolayer},\ }\href
  {https://doi.org/10.1103/PhysRevMaterials.3.044001} {\bibfield  {journal}
  {\bibinfo  {journal} {Phys. Rev. Mater.}\ }\textbf {\bibinfo {volume} {3}},\
  \bibinfo {pages} {044001} (\bibinfo {year} {2019})}\BibitemShut {NoStop}%
\bibitem [{\citenamefont {Grimme}\ \emph {et~al.}(2010)\citenamefont {Grimme},
  \citenamefont {Antony}, \citenamefont {Ehrlich},\ and\ \citenamefont
  {Krieg}}]{DFT-D3}%
  \BibitemOpen
  \bibfield  {author} {\bibinfo {author} {\bibfnamefont {S.}~\bibnamefont
  {Grimme}}, \bibinfo {author} {\bibfnamefont {J.}~\bibnamefont {Antony}},
  \bibinfo {author} {\bibfnamefont {S.}~\bibnamefont {Ehrlich}},\ and\ \bibinfo
  {author} {\bibfnamefont {H.}~\bibnamefont {Krieg}},\ }\bibfield  {title}
  {\bibinfo {title} {A consistent and accurate ab initio parametrization of
  density functional dispersion correction ($\mathrm{DFT}$-$\mathrm{D}$) for
  the 94 elements $\mathrm{H}$-$\mathrm{Pu}$},\ }\href
  {https://doi.org/10.1063/1.3382344} {\bibfield  {journal} {\bibinfo
  {journal} {J. Chem. Phys.}\ }\textbf {\bibinfo {volume} {132}},\ \bibinfo
  {pages} {154104} (\bibinfo {year} {2010})}\BibitemShut {NoStop}%
\bibitem [{\citenamefont {King-Smith}\ and\ \citenamefont
  {Vanderbilt}(1993)}]{BF}%
  \BibitemOpen
  \bibfield  {author} {\bibinfo {author} {\bibfnamefont {R.~D.}\ \bibnamefont
  {King-Smith}}\ and\ \bibinfo {author} {\bibfnamefont {D.}~\bibnamefont
  {Vanderbilt}},\ }\bibfield  {title} {\bibinfo {title} {Theory of polarization
  of crystalline solids},\ }\href {https://doi.org/10.1103/PhysRevB.47.1651}
  {\bibfield  {journal} {\bibinfo  {journal} {Phys. Rev. B}\ }\textbf {\bibinfo
  {volume} {47}},\ \bibinfo {pages} {1651} (\bibinfo {year}
  {1993})}\BibitemShut {NoStop}%
\bibitem [{\citenamefont {Lou}\ \emph {et~al.}(2021)\citenamefont {Lou},
  \citenamefont {Li}, \citenamefont {Ji}, \citenamefont {Yu}, \citenamefont
  {Feng}, \citenamefont {Gong},\ and\ \citenamefont {Xiang}}]{PASP}%
  \BibitemOpen
  \bibfield  {author} {\bibinfo {author} {\bibfnamefont {F.}~\bibnamefont
  {Lou}}, \bibinfo {author} {\bibfnamefont {X.}~\bibnamefont {Li}}, \bibinfo
  {author} {\bibfnamefont {J.}~\bibnamefont {Ji}}, \bibinfo {author}
  {\bibfnamefont {H.}~\bibnamefont {Yu}}, \bibinfo {author} {\bibfnamefont
  {J.}~\bibnamefont {Feng}}, \bibinfo {author} {\bibfnamefont {X.}~\bibnamefont
  {Gong}},\ and\ \bibinfo {author} {\bibfnamefont {H.}~\bibnamefont {Xiang}},\
  }\bibfield  {title} {\bibinfo {title} {$\mathrm{PASP}$: Property analysis and
  simulation package for materials},\ }\href
  {https://doi.org/10.1063/5.0043703} {\bibfield  {journal} {\bibinfo
  {journal} {J. Chem. Phys.}\ }\textbf {\bibinfo {volume} {154}},\ \bibinfo
  {pages} {114103} (\bibinfo {year} {2021})}\BibitemShut {NoStop}%
\bibitem [{\citenamefont {Pan}\ \emph {et~al.}(2024)\citenamefont {Pan},
  \citenamefont {Xu}, \citenamefont {Li}, \citenamefont {Xu}, \citenamefont
  {Liu}, \citenamefont {Gu},\ and\ \citenamefont {Duan}}]{Weiyi2}%
  \BibitemOpen
  \bibfield  {author} {\bibinfo {author} {\bibfnamefont {W.}~\bibnamefont
  {Pan}}, \bibinfo {author} {\bibfnamefont {C.}~\bibnamefont {Xu}}, \bibinfo
  {author} {\bibfnamefont {X.}~\bibnamefont {Li}}, \bibinfo {author}
  {\bibfnamefont {Z.}~\bibnamefont {Xu}}, \bibinfo {author} {\bibfnamefont
  {B.}~\bibnamefont {Liu}}, \bibinfo {author} {\bibfnamefont {B.-L.}\
  \bibnamefont {Gu}},\ and\ \bibinfo {author} {\bibfnamefont {W.}~\bibnamefont
  {Duan}},\ }\bibfield  {title} {\bibinfo {title} {Chiral magnetism in
  lithium-decorated monolayer $\mathrm{CrTe}_{2}$: Interplay between
  dzyaloshinskii-moriya interaction and higher-order interactions},\ }\href
  {https://doi.org/10.1103/PhysRevB.109.214405} {\bibfield  {journal} {\bibinfo
   {journal} {Phys. Rev. B}\ }\textbf {\bibinfo {volume} {109}},\ \bibinfo
  {pages} {214405} (\bibinfo {year} {2024})}\BibitemShut {NoStop}%
\bibitem [{\citenamefont {Xu}\ \emph {et~al.}(2022)\citenamefont {Xu},
  \citenamefont {Li}, \citenamefont {Chen}, \citenamefont {Zhang},
  \citenamefont {Xiang},\ and\ \citenamefont {Bellaiche}}]{XCS2}%
  \BibitemOpen
  \bibfield  {author} {\bibinfo {author} {\bibfnamefont {C.}~\bibnamefont
  {Xu}}, \bibinfo {author} {\bibfnamefont {X.}~\bibnamefont {Li}}, \bibinfo
  {author} {\bibfnamefont {P.}~\bibnamefont {Chen}}, \bibinfo {author}
  {\bibfnamefont {Y.}~\bibnamefont {Zhang}}, \bibinfo {author} {\bibfnamefont
  {H.}~\bibnamefont {Xiang}},\ and\ \bibinfo {author} {\bibfnamefont
  {L.}~\bibnamefont {Bellaiche}},\ }\bibfield  {title} {\bibinfo {title}
  {Assembling diverse skyrmionic phases in
  $\mathrm{Fe}$$_{3}$$\mathrm{GeTe}$$_{2}$ monolayers},\ }\href
  {https://onlinelibrary.wiley.com/doi/full/10.1002/adma.202107779} {\bibfield
  {journal} {\bibinfo  {journal} {Adv. Mater.}\ }\textbf {\bibinfo {volume}
  {34}},\ \bibinfo {pages} {2107779} (\bibinfo {year} {2022})}\BibitemShut
  {NoStop}%
\bibitem [{\citenamefont {Ni}\ \emph {et~al.}(2021)\citenamefont {Ni},
  \citenamefont {Li}, \citenamefont {Amoroso}, \citenamefont {He},
  \citenamefont {Feng}, \citenamefont {Kan}, \citenamefont {Picozzi},\ and\
  \citenamefont {Xiang}}]{NiI2Xiang}%
  \BibitemOpen
  \bibfield  {author} {\bibinfo {author} {\bibfnamefont {J.~Y.}\ \bibnamefont
  {Ni}}, \bibinfo {author} {\bibfnamefont {X.~Y.}\ \bibnamefont {Li}}, \bibinfo
  {author} {\bibfnamefont {D.}~\bibnamefont {Amoroso}}, \bibinfo {author}
  {\bibfnamefont {X.}~\bibnamefont {He}}, \bibinfo {author} {\bibfnamefont
  {J.~S.}\ \bibnamefont {Feng}}, \bibinfo {author} {\bibfnamefont {E.~J.}\
  \bibnamefont {Kan}}, \bibinfo {author} {\bibfnamefont {S.}~\bibnamefont
  {Picozzi}},\ and\ \bibinfo {author} {\bibfnamefont {H.~J.}\ \bibnamefont
  {Xiang}},\ }\bibfield  {title} {\bibinfo {title} {Giant biquadratic exchange
  in 2$\mathrm{D}$ magnets and its role in stabilizing ferromagnetism of
  $\mathrm{NiCl}_{2}$ monolayers},\ }\href
  {https://doi.org/10.1103/PhysRevLett.127.247204} {\bibfield  {journal}
  {\bibinfo  {journal} {Phys. Rev. Lett.}\ }\textbf {\bibinfo {volume} {127}},\
  \bibinfo {pages} {247204} (\bibinfo {year} {2021})}\BibitemShut {NoStop}%
\bibitem [{\citenamefont {Li}\ \emph {et~al.}(2020)\citenamefont {Li},
  \citenamefont {Lou}, \citenamefont {Gong},\ and\ \citenamefont
  {Xiang}}]{LXY}%
  \BibitemOpen
  \bibfield  {author} {\bibinfo {author} {\bibfnamefont {X.-Y.}\ \bibnamefont
  {Li}}, \bibinfo {author} {\bibfnamefont {F.}~\bibnamefont {Lou}}, \bibinfo
  {author} {\bibfnamefont {X.-G.}\ \bibnamefont {Gong}},\ and\ \bibinfo
  {author} {\bibfnamefont {H.}~\bibnamefont {Xiang}},\ }\bibfield  {title}
  {\bibinfo {title} {Constructing realistic effective spin hamiltonians with
  machine learning approaches},\ }\href
  {https://iopscience.iop.org/article/10.1088/1367-2630/ab85df} {\bibfield
  {journal} {\bibinfo  {journal} {New J. Phys.}\ }\textbf {\bibinfo {volume}
  {22}},\ \bibinfo {pages} {053036} (\bibinfo {year} {2020})}\BibitemShut
  {NoStop}%
\bibitem [{\citenamefont {Hukushima}\ and\ \citenamefont
  {Nemoto}(1996)}]{PTMC}%
  \BibitemOpen
  \bibfield  {author} {\bibinfo {author} {\bibfnamefont {K.}~\bibnamefont
  {Hukushima}}\ and\ \bibinfo {author} {\bibfnamefont {K.}~\bibnamefont
  {Nemoto}},\ }\bibfield  {title} {\bibinfo {title} {Exchange monte carlo
  method and application to spin glass simulations},\ }\href
  {https://journals.jps.jp/doi/abs/10.1143/JPSJ.65.1604} {\bibfield  {journal}
  {\bibinfo  {journal} {J. Phys. Soc. Jpn}\ }\textbf {\bibinfo {volume} {65}},\
  \bibinfo {pages} {1604} (\bibinfo {year} {1996})}\BibitemShut {NoStop}%
\bibitem [{\citenamefont {Stiefel}(1952)}]{CG}%
  \BibitemOpen
  \bibfield  {author} {\bibinfo {author} {\bibfnamefont {E.}~\bibnamefont
  {Stiefel}},\ }\bibfield  {title} {\bibinfo {title} {Methods of conjugate
  gradients for solving linear systems},\ }\href@noop {} {\bibfield  {journal}
  {\bibinfo  {journal} {J. Res. Nat. Bur. Standards}\ }\textbf {\bibinfo
  {volume} {49}},\ \bibinfo {pages} {409} (\bibinfo {year} {1952})}\BibitemShut
  {NoStop}%
\bibitem [{\citenamefont {Amoroso}\ \emph {et~al.}(2020)\citenamefont
  {Amoroso}, \citenamefont {Barone},\ and\ \citenamefont
  {Picozzi}}]{NiI2Silvia}%
  \BibitemOpen
  \bibfield  {author} {\bibinfo {author} {\bibfnamefont {D.}~\bibnamefont
  {Amoroso}}, \bibinfo {author} {\bibfnamefont {P.}~\bibnamefont {Barone}},\
  and\ \bibinfo {author} {\bibfnamefont {S.}~\bibnamefont {Picozzi}},\
  }\bibfield  {title} {\bibinfo {title} {Spontaneous skyrmionic lattice from
  anisotropic symmetric exchange in a $\mathrm{Ni}$-halide monolayer},\ }\href
  {https://www.nature.com/articles/s41467-020-19535-w} {\bibfield  {journal}
  {\bibinfo  {journal} {Nat. Commun.}\ }\textbf {\bibinfo {volume} {11}},\
  \bibinfo {pages} {5784} (\bibinfo {year} {2020})}\BibitemShut {NoStop}%
\end{thebibliography}
\end{document}